\documentclass[
	aps, prd,  notitlepage, 
        floats, floatfix, onecolumn,
	amsmath, amssymb, amsfonts, eqsecnum,
	superscriptaddress,
	showpacs, showkeys,
	nofootinbib,
 	longbibliography,
]{revtex4-1}
\usepackage{multirow}
\usepackage{graphicx}
\usepackage{subfigure}
\usepackage[usenames,dvipsnames]{xcolor}
\usepackage{tikz}
\usepackage{xspace} 
\usepackage{bm}
\usepackage[utf8]{inputenc} 
\usepackage{latexsym}
\usepackage{float}
\usepackage{ulem}
\renewcommand{\emph}[1]{\textit{#1}}
\xdefinecolor{mylinkcolor}{rgb}{0,0,0.5}
\usepackage[
	bookmarksnumbered, bookmarksopen, bookmarksopenlevel=2,
	breaklinks=true,
	colorlinks=true, filecolor=mylinkcolor, citecolor=mylinkcolor,
	linkcolor=mylinkcolor, urlcolor=mylinkcolor, menucolor=mylinkcolor,
]{hyperref}

\def\be{\begin{equation}}
\def\ee{\end{equation}}
\def\bea{\begin{eqnarray}}
\def\eea{\end{eqnarray}}
\newcommand{\bes}{\begin{subequations}}
\newcommand{\ees}{\end{subequations}}
\def\comment#1{}

\usepackage{color}

\allowdisplaybreaks

\begin{document}
 
\bibliographystyle{unsrt}
\title{Surrogate model for gravitational waveforms of spin-aligned binary black holes with eccentricities}

\author{Qianyun Yun}

\affiliation{Shanghai Astronomical Observatory,  Chinese Academy of Sciences,  Shanghai,  China,  200030}
\affiliation{School of Physical Science and Technology, ShanghaiTech University, Shanghai,  China, 201210}
\affiliation{School of Astronomy and Space Science,  University of Chinese Academy of Sciences,  Beijing,  China,  100049}

\author{Wen-Biao Han}
\email{Conrresponding author: wbhan@shao.ac.cn}

\affiliation{Shanghai Astronomical Observatory,  Chinese Academy of Sciences,  Shanghai,  China,  200030}
\affiliation{School of Fundamental Physics and Mathematical Sciences, Hangzhou Institute for Advanced Study, UCAS, Hangzhou 310024, China}
\affiliation{School of Astronomy and Space Science,  University of Chinese Academy of Sciences,  Beijing,  China,  100049}

\author{Xingyu Zhong}

\affiliation{Shanghai Astronomical Observatory,  Chinese Academy of Sciences,  Shanghai,  China,  200030}

\affiliation{School of Astronomy and Space Science,  University of Chinese Academy of Sciences,  Beijing,  China,  100049}

\author{Carlos~A.~Benavides-Gallego}
\affiliation{Shanghai Astronomical Observatory,  Chinese Academy of Sciences,  Shanghai,  China,  200030}


\begin{abstract}
A waveform model for the eccentric binary black holes named SEOBNRE has been used to analyze the LIGO-Virgo's gravitational wave data by several groups. The accuracy of this model has been validated by comparing it with numerical relativity. However, SEOBNRE is a time-domain model, and the efficiency for generating waveforms is a bottleneck in data analysis. To overcome this disadvantage, we offer a reduced-order surrogate model for eccentric binary black holes based on the SEOBNRE waveforms.
This surrogate model (SEOBNRE\_S) can simulate the complete inspiral-merger-ringdown waves with enough accuracy, covering eccentricities from 0 to 0.25 (0.1), and mass ratio from 1:1 to 5:1 (2:1) for nonspinning (spinning) binaries. The speed of waveform generation is accelerated about $10^2 \sim 10^3$ times than the original SEOBNRE model. Therefore SEOBNRE\_S could be helpful in the analysis of LIGO data to find potential eccentricities.  \\

\end{abstract}
\maketitle
\section{Introduction}
The direct detection of gravitational waves from the mergers of compact binaries has opened a new era for gravitational wave (GW) astronomy~\cite{abbott2016tests}.
Recently, the catalog GWTC-2 has updated all the events observed from the first, second and third runs~\cite{abbott2019gwtc,abbott2020gwtc} (O1, O2 and O3a, respectively) of the Advanced LIGO~\cite{TheLIGOScientific:2014jea,abbott2018prospects} and 
Advanced Virgo~\cite{TheVirgo:2014hva}.
Moreover, the catalog has listed 44 credible binary black hole (BBH) events and two binary neutron star events~\cite{TheLIGOScientific:2017qsa,Abbott:2020uma}. In all these observations, the number of events observed in O3a is more than twice of O1 and O2. Therefore, as more and more gravitational wave events are detected, we will have the opportunity to understand how black hole binary systems form in the Universe.  

Although these detections have already provided plenty of information on merger processes, the formation mechanism of binaries systems is still an open issue in astrophysics~\cite{Linden:2010jx,Ivanova:2014tpa,Zuo:2014qra,Rodriguez:2016kxx,Mandel:2015qlu,Rodriguez:2016avt}. To explain the formation of binaries, the community has proposed several formation channels, which involve specific environments and physical processes \cite{TheLIGOScientific:2016htt}.
However, it is believed that two canonical formation channels exist for compact binaries: isolated binary evolution ~\cite{bethe1998evolution,belczynski2002comprehensive,belczynski2014formation,belczynski2016first,spera2015mass} and dynamical formation~\cite{zwart1999black,o2006binary,sadowski2008total,downing2010compact,downing2011compact}.  In the first channel, for example, a binary system can be formed from the evolution of a common-envelope phase~\cite{Bethe:1998bn,PortegiesZwart1:1997zn,Belczynski:2001uc,Dominik:2014yma}, from the remnants of population III~\cite{Mandel:2015qlu,Inayoshi:2017mrs}, or through chemically homogeneous evolution of close binaries that attain rapid rotation~\cite{marchant2016new,deMink:2016vkw}. It is well known that isolated BBHs typically process little eccentricity (ecc)~\cite{Randall:2019znp}, and the gravitational radiation will circularize binary orbits~\cite{Randall:2017jop}. One can see Ref.~\cite{deMink:2016vkw} and references therein for more details.

On the other hand, eccentricity can survive if there is some other source that drives an increase during inspiral~\cite{Randall:2017jop,Rodriguez:2018pss}; these possibilities include three-body systems ~\cite{Zevin:2017evb,OLeary:2005vqo,Gultekin:2005fd,Silsbee:2016djf}, and dense environments with many BHs created~\cite{OLeary:2008myb}. Hence, the formation of binaries takes place in dense stellar environments such as young stellar clusters~\cite{rodriguez2015binary,haster2016n,chatterjee2017dynamical,banerjee2017stellar}, globular clusters(GCs), galactic nuclei~\cite{Kulkarni:1993fr,Sigurdsson:1993zrm,zwart1999black}, or three-body systems~\cite{Antonini:2013tea,Kimpson:2016dgk,Antonini:2017ash}. In this dense stellar environment, BBHs are created and harden through dynamical interactions~\cite{Quinlan:1996vp,Rodriguez:2018pss}. Moreover, some researchers also predicted that about 5\% of dynamically formed binaries have $ecc \geq 0.1$ at 10 Hz~\cite{samsing2018eccentric,rodriguez2018post}.

In addition to these canonical scenarios, the famous Kozai-Lidov (KL) mechanism happens in a triple system where the BBH is the inner binary. Therefore the KL mechanism can affect the merge of the binary system and trigger the oscillations of the BBH's eccentricity~\cite{Silsbee:2016djf,Hoang:2017fvh,Gupta:2019unn,Wen:2002km,VanLandingham:2016ccd,Hoang:2017fvh}. Recently, Ref.~\cite{takatsy2019eccentricity} claimed that the ratio of eccentric binaries in the LIGO band is 10~\% for binaries formed through the Kozai-Lidov mechanism, and it could be as large as 90~\% for gravitational capture formation. For a comprehensive review for all these BBH formation channels, please read Ref.~\cite{Gerosa:2021mno}.

Therefore, BBHs detected by the advanced LIGO/Virgo usually form from the isolated or dynamical channels. Nevertheless, the task of determining which is the formation channel represents a challenge. One possibility to overcome this difficulty is to analyze the orbit of the binary system. In particular, since the eccentricity  is a prominent feature of BBHs, one can use it to infer the formation channels. It is well known that the orbit becomes circular at the last stage of the merger~\cite{peters1964gravitational,hinder2008circularization} due to gravitational radiation. However, at the earlier stage (when the frequency of the GW is about 10~Hz), the remnants of eccentricities of different formation channels could be different too. For example, if the merger comes from the isolated binary evolution channel, the eccentricity  is negligible~\cite{peters1964gravitational,hinder2008circularization}. Supposing that the merger comes from the dynamical channel, the eccentricity is noticeable at 10~Hz~\cite{zevin2017constraining,zevin2019eccentric,zevin2019can,rodriguez2018triple,samsing2018eccentric,gondan2019measurement,samsing2018eccentric,rodriguez2018post,rodriguez2018post,takatsy2019eccentricity}. Hence, due to the unceasing update of LIGO, the gravitational wave signal at 10 Hz with lower noise will be available in the future, giving us the possibility to investigate formation channels. For example, Ref.~\cite{Gayathri:2020coq} has claimed that GW190521 is a highly eccentric BBH merger, and this implies that GW190521-like binaries may form dynamically\cite{Holgado:2020imj}. In this sense, searching eccentric binaries is interesting in GW astronomy~\cite{lower2018measuring,romero2019searching,Salemi:2019owp,nitz2020search}.

Parameter estimations for eccentric sources in GW
data need exact waveform templates. Nevertheless,
the orbital evolution and waveform of an elliptic orbit are more complicated than a circular orbit due to the eccentricity. In the literature, it is possible to find several waveform templates that consider the eccentricity of the orbit~\cite{vallisneri2015ligo,Huerta:2016rwp,East:2012xq,cao2017waveform,Hinderer:2017jcs,moore20193pn,huerta2014accurate,tiwari2016proposed,Yun_2020,Huerta:2017kez}. First, in LIGO's LALSuite library\footnote{https://wiki.ligo.org/DASWG/LALSuite}\cite{vallisneri2015ligo}, we find some templates  with eccentricities such as Taylor F2\cite{moore20193pn}, eccTD and eccFD\cite{huerta2014accurate,tiwari2016proposed}. Since these templates only contain the inspiral part, they will lose signal-noise-ratio and the accuracy is not good enough for data analysis~\cite{Yun_2020}.  Another eccentric waveform template~\cite{Huerta:2017kez} considered an eccentric, nonspinning approximate model called ENIGMA with mass ratios up to 5.5 and $ecc \lesssim 0.2$. 
Moreover, the so-called SEOBNRE model~\cite{cao2017waveform} can include eccentricities by combining the quasicircular dynamics of effective-one-body-numerical-relativity (SEOBNR)~\cite{buonanno1999effective,Taracchini:2012ig,barausse2011extending,taracchini2012prototype,lovelace2016modeling,cao2017waveform} and eccentric post-Newtonian (PN) corrections (up to 2PN order). The overlaps between the SEOBNRE model and the numerical relativity model are better than 99\% for the mass ratio, spin and eccentricity in a range of [1,10], [0,0.5] and [0,0.3] respectively. This means that SEOBNRE is accurate enough to analyze  the  LIGO-Virgo  GW  data and several groups have used this model to search for eccentricities in the GW events, see Refs.~\cite{lower2018measuring, Romero-Shaw:2020thy} and references therein.

 
However, although the accuracy of the SEOBNRE model is enough for data analysis, the computation speed makes the model difficult for parameter estimations because it is a time-domain evolution waveform. On the other hand, a parameter estimation of the GW source  needs to evaluate millions of waveforms while generating SEOBNRE waveforms of a single BBH merger usually needs about 10 CPU seconds. Therefore, it is necessary to build waveform models that include eccentricity not only accurate but also efficient enough to implement data analysis.

Therefore, fast surrogate waveform models\cite{Field:2013cfa,Blackman:2015pia,Blackman:2017dfb,Blackman:2017pcm,Varma:2018aht,Varma:2018mmi,Varma:2019csw} are now often used in
LIGO's data processing pipeline. For example, the NRSur7dq4 model~\cite{Varma:2019csw} has been used to estimate the parameters of GW190521 ~\cite{Abbott:2020tfl}. This kind of numerical relativity(NR) surrogate model was constructed with the help of an algorithm provided by~\cite{Field:2013cfa}. This algorithm generates accurate surrogate models without considering physics inside. \comment{According to Ref.~\cite{Field:2013cfa}, several works have shown that gravitational waveforms evince a redundancy in the parameter space. As a consequence, one can use a smaller amount of information to represent a fiducial waveform. Hence, most of these surrogate models use a few number of representative waveforms, which can be selected by a greedy algorithm~\cite{Field:2013cfa}. Reduced basis (RB) is the base thought of building those surrogate models.} On the other hand, there are some EOB surrogate models~\cite{Purrer:2014fza,Purrer:2015tud,Nagar:2018zoe,Cotesta:2020qhw}. These works employed singular value decomposition as a new way to construct the surrogate model. Nevertheless, there is still no surrogate waveform model, whose intrinsic parameters cover varied mass ratios, orbital eccentricity and the black hole spin.

In this paper, we first propose a surrogate model for spinning BBHs with varied mass ratios, and orbital eccentricity based on the Python Package Rompy\footnote{This was imported from https://bitbucket.org/chadgalley/rompy}. To describe the binary black hole system, we use the following parameters. The masses of two black holes are $m_1$
and $m_2$, with $m_1 \geq m_2$.  The total mass is
$M = m_1 + m_2$, and the mass ratio is $q \equiv m_1/m_2$ ($q $ is always larger than $1$), and $S_1$ and $S_2$ are used to denote the spin of two black holes. The dimensionless spins are then $\chi_1=S_1/m_1^2$ , $\chi_2=S_2/m_2^2$ and the effective spin $\chi_{\rm eff}=(m_1\chi_1+m_2\chi_2)/M$. In our model, the mass ratio $q$ is from 1 to 5, the eccentricity is up to 0.25 and spin is in a range of -0.5 to 0.5.

This paper is organized as follows. In Sec.~II, we briefly introduce how to build a surrogate model, then in Sec~III, we evaluate the SEOBNRE surrogate by validating the surrogate waveforms and calculating the generating time. Finally, a discussion will be addressed in Sec.~IV.

\section{Surrogate waveform model}

In this section, we first explain why it is necessary to build a fast surrogate model SEOBNRE\_S with eccentricities. Secondly, we discuss the process of constructing surrogate models.
\subsection{Problem statement}
There are two main steps in the detection of gravitational waves: 1. find the GW signal in detector data; 2. infer the physical parameters of the source according to the GW signal. When we search signals, a prevailing technique for detecting a compact binary coalescence is to match filter the detector data onto the data template banks~\cite{Allen:2005fk}. Therefore, it is necessary to construct a template bank in advance. After a gravitational wave signal is detected, parameter estimation (PE) algorithms can estimate the posterior probabilities of the binary system's parameters, and help to explore the underlying physics of the compact objects.
Both steps  need to generate a large number of accurate waveforms for data analysis. In this sense,  if we want to study the orbital eccentricity of double black holes, we need to build a model that not only considers eccentricity but can also quickly and accurately generate waveforms.
\begin{figure}
  \centering 
  \includegraphics[height=2.4in]{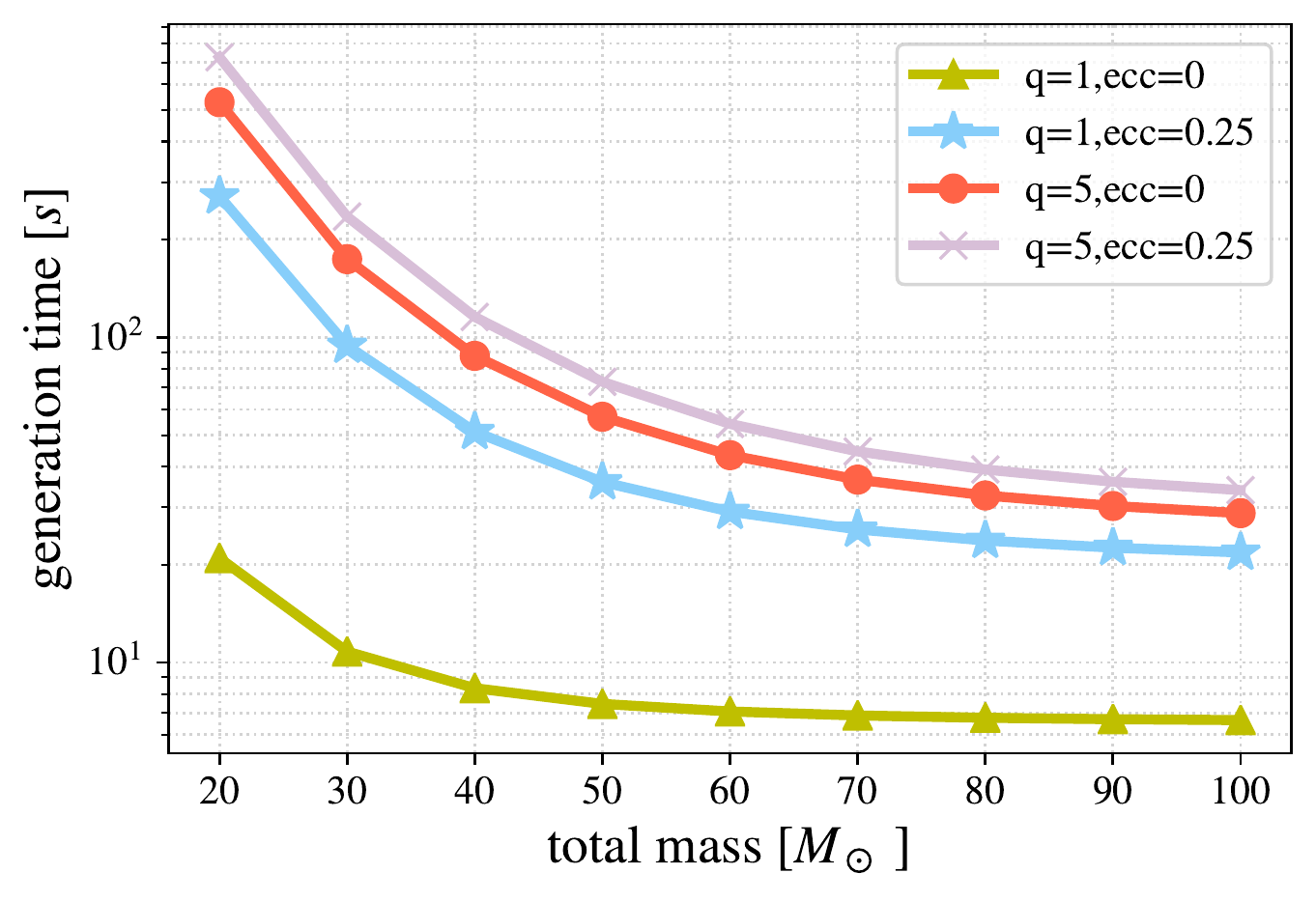}
  \caption{The generation time for a time-domain SEOBNRE waveform as a function of the total mass $M$ for different values of $q$ and $ecc$. The starting frequency is 10~Hz, the sampling frequency is 4069~Hz, and $\chi=0$.}
  \label{generate time}
  
\end{figure}

 Previous works have validated SEOBNRE waveforms with SXS\footnote{SXS is a numerical relativity waveform that can be download at: https://www.black-holes.org/code/SpEC.html.} waveforms~\cite{liu2019validating,Yun_2020}, where the SEOBNRE model has been proved very accurate. Nevertheless, it is difficult to use the SEOBNRE model directly for data analysis due to the numerical calculations of a set of ordinary differential equations~\cite{cao2017waveform}. This makes the generation of SEOBNRE waveforms is slow (see Fig.~\ref{generate time}).

From the figure, we can see that the generation time of a time-domain SEOBNRE waveform varies from a few seconds to a few hundred seconds by using one single CPU core [Intel(R) CPU E5-2687W 0 @ 3.10~Ghz].  For binary black holes with the equal masses ($q=1$) and zero eccentricity ($ecc=0$), the generation time goes from 5 to 20 seconds (see the solid-green line). When we increase the mass ratio, the generation time also increases because $q$ affects the inspiral time of the binary(see the solid-red line).   For example, a waveform with $q=5$ and total mass $20M$ takes about 700 seconds to generate, while a waveform with $q=1$ takes around 20 seconds. For the smaller total mass, the generation time of the waveform will be longer. The reason for this behavior lies in the fact that a BBH with smaller $M$ has more cycles before the merger than the one with larger $M$ (if we set their starting frequency at the same value). Therefore, the generation time decreases monotonically with the total mass. On the other hand, when we consider the eccentricity, the figure shows that the generation time is longer, too. For example, for $ecc = 0.25$, $q=1$ and $M=20M_\odot$, the generation time of a waveform is about 400 seconds, which is much longer than the non eccentricity case. When the mass ratio increases, the generation time of the waveform also increases: a  waveform with $q=5$, $ecc=0.25$, and a total mass of 20$M_\odot$ takes about 800 seconds to generate (see the purple line).

As shown in Fig.~\ref{generate time}, the long generation time means that it is CPU expensive to use SEOBNRE directly for the construction of template banks and PE. It is necessary to generate $10^6-10^7$ waveforms in 11-dimensional parameter space in PE. (without eccentricity)~\cite{Aasi:2014tra}. On the other hand, if we want to analyze an elliptical binary system, the number of waveforms needed will increase. For example, if we assume the generation time of a waveform 
is 10 seconds,
it will take several months to do parameter estimation with only one CPU core. If the generation time of a single waveform is longer, such as 700 seconds (see Fig.~\ref{generate time} when the mass ratio $q=5$ and total mass $M=20M_\odot$), it will take more than 10 years to complete the PE. In this sense, it is necessary to establish a fast surrogate
model based on the SEOBNRE for
parameter estimations of eccentric BBHs.


\subsection{Training points selection}

 It is known that  the gravitational waves can be expressed in terms of its two polarizations $h_{+}$ and $h_{\times}$,
 \begin{equation}
     h(t, \theta, \phi ; \lambda)=h_{+}(t, \theta, \phi ; \lambda)-i h_{\times}(t, \theta, \phi ; \lambda)\,
 \end{equation}
 
where $t$ denotes the time and $\theta$, $\phi$ denote the polar and azimuthal angles of the source, respectively. In this paper, $\lambda$ represents a set of parameters that characterize the waveform: $m_{1},m_{2}$, $ecc$ and $\chi_1,\chi_2$. In this work, we only consider $(l,m)=(2,2)$  mode.
 
 
The eccentricity is less than 0.3, most of the SEOBNRE waveforms can match with the SXS waveforms better than 99\% ~\cite{liu2019validating,Yun_2020}. Therefore, we also consider eccentricities smaller than 0.3. The mass ratio $q $ is limited up to 5 (which is suitable for most signals that LIGO-Virgo has detected) and since the accuracy of SEOBNRE is largely affected by the spin hang-up effect, the spin of each black hole is from $-0.5$ to $0.5$.
\begin{figure}
  \centering 
  \includegraphics[height=2.3in]{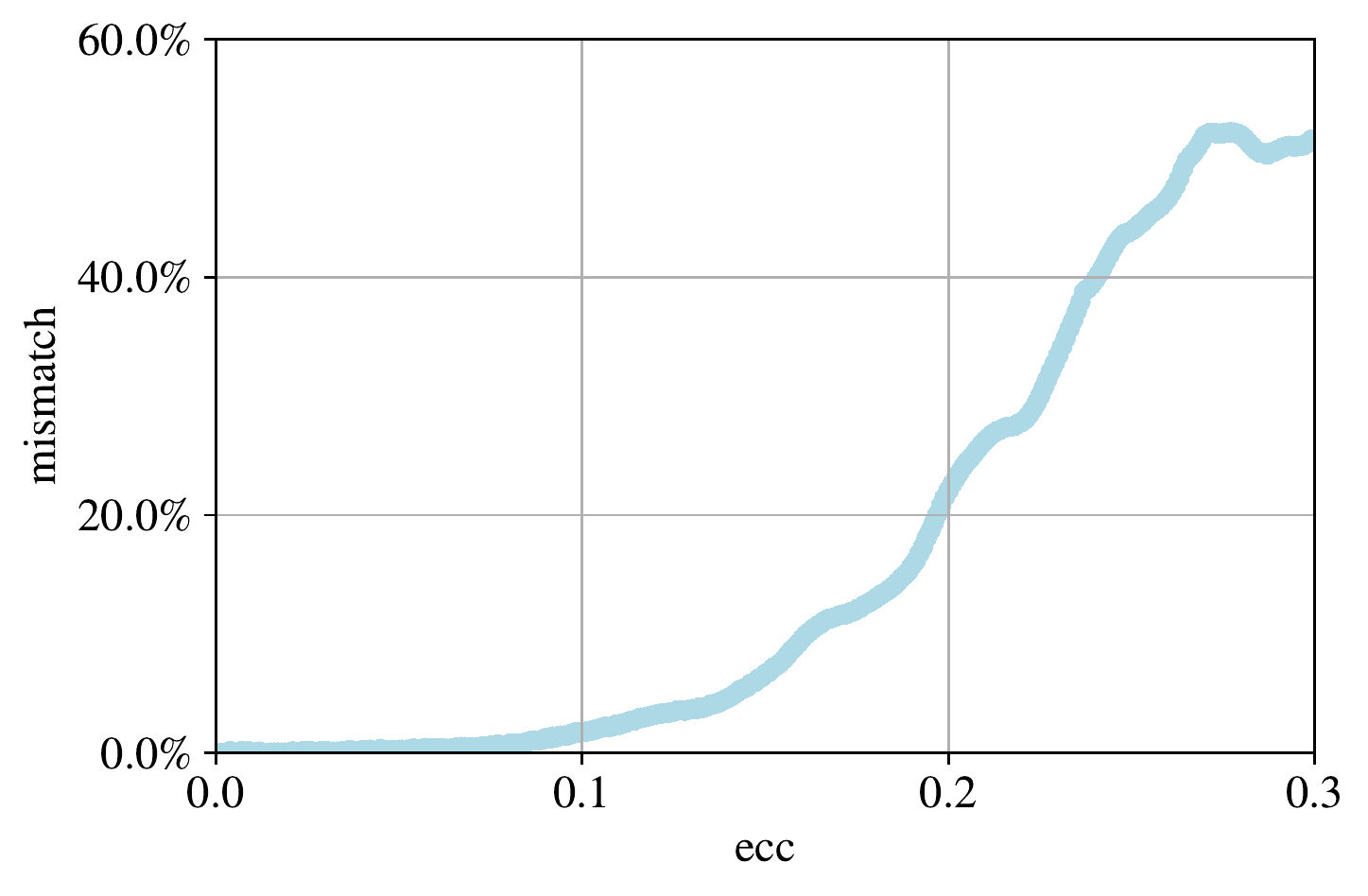}
  \caption{The mismatch between two waveforms $h\left(\lambda;ecc_0\right)$ and $h\left(\lambda;ecc\right)$. Parameters are set as $\chi_1=\chi_2=0$, $m_1=m_2=20M_\odot$, $ecc_0=0$ and the initial frequency is $10~\text{Hz}$.}
  \label{dis}
  
\end{figure}

According to Fig.~\ref{dis}, the mismatch [see Eq.(~\ref{mismatcheq})] between the gravitational waveforms $h\left(\lambda;ecc_0\right)$ and $h\left(\lambda;ecc\right)$ changes dramatically  when the eccentricity increases. Therefore, it is difficult to build a surrogate model covering the whole parameter space. For this reason, to avoid problems in the accuracy and computational cost,  we generated piecewise models to cover the parameter space. In this way, the eccentricity parameters are divided into several segments and each segment will generate its own surrogate model. Therefore, based on the input eccentricity,  waveforms are generated by the corresponding surrogate model. This kind of piecewise models not only reduces the computing time but also makes our surrogate waveform more accurate.

In this way, we choose the training data for the nonspinning case. The eccentricity is divided into seven segments: [0,0.1], [0.1,0.15], [0.15,0.20], [0.2,0.225], [0.225,0.25], [0.25,0.275] and [0.275,0.3]. It is important to point out that the eccentricity in each segment is sampled by a different $\Delta ecc$, i.e., $\Delta ecc =0.002$ in the first segment ([0,0.1]), $\Delta ecc=0.001$ in the second and third segments, $\Delta ecc=0.0005$ in the fourth and fifth segments ([0.2,0.225] and [0.225,0.25]), and $\Delta ecc=0.0002$ in the last two segments. The number of training data for each segment is shown in Table.~\ref{ nonspinning surrogate training space}. It needs to take five days to generate the nonspinning SEOBNRE waveforms for the first five segments and four days for the last two.

\begin{figure}
  \centering 
  { 
    \includegraphics[height=2.0in]{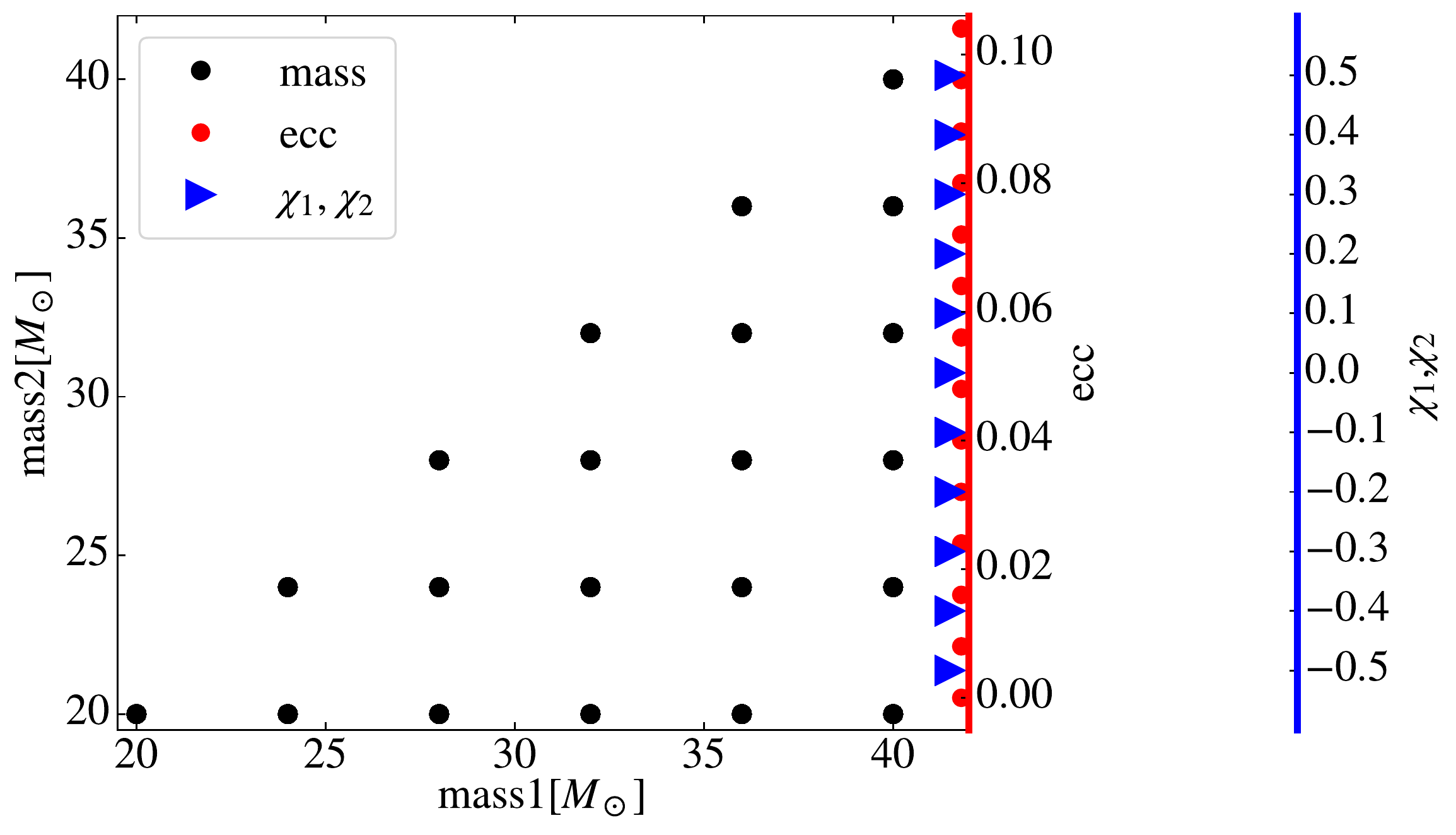} 

  } 
    
  \caption{The training data for the spin-aligned surrogate model. As same as Fig~\ref{Tdatanonspinning}, black dots represent the masses of the two black holes, while the red dots represent how we sample the eccentricity. The blue triangles represent the sampling of the dimensionless spins, corresponding to the blue axis on the left side of the figure.
} 
  \label{training space with spin} 
\end{figure}

\begin{table}[htb]
\centering
  \begin{tabular}{|c|c|}
  \hline\hline
   ecc range &Training  data\\
  \hline
  $0-0.1$ &$3366$\\
  
  $0.1-0.15$&$3366$  \\
  
  $0.15-0.2$&$3366$\\
  
  $0.2-0.225$&$3366$  \\
  
  $0.225-0.25$&$3366$  \\
  $0.25-0.275$&$6666$  \\
  $0.275-0.30$&$6666$  \\

  \hline\hline
  \end{tabular}
  \caption{The number of SEOBNRE waveforms in each segment. A total of 26796 SEOBNRE waveforms are used to construct the nonspinning surrogate model.}  
  \label{ nonspinning surrogate training space}
	

\end{table}

The training data for the spin-aligned surrogate model is shown in Fig.~\ref{training space with spin}. We consider $q\in[1,2]$, $ecc\in[0,0.1] $, 
 $\chi_1, \chi_2 \in[-0.5,0.5]$, $\Delta ecc = 0.008$, and $\Delta \chi_1=\Delta \chi_2= 0.01$. 35565 training data are used to built the spin-aligned surrogate model. In this case, the generation time of the spin-aligned SEOBNRE waveforms is 12 days.

\subsection{Building the model}
\begin{figure}[htb]
  \centering 
  \includegraphics[height=2.1in]{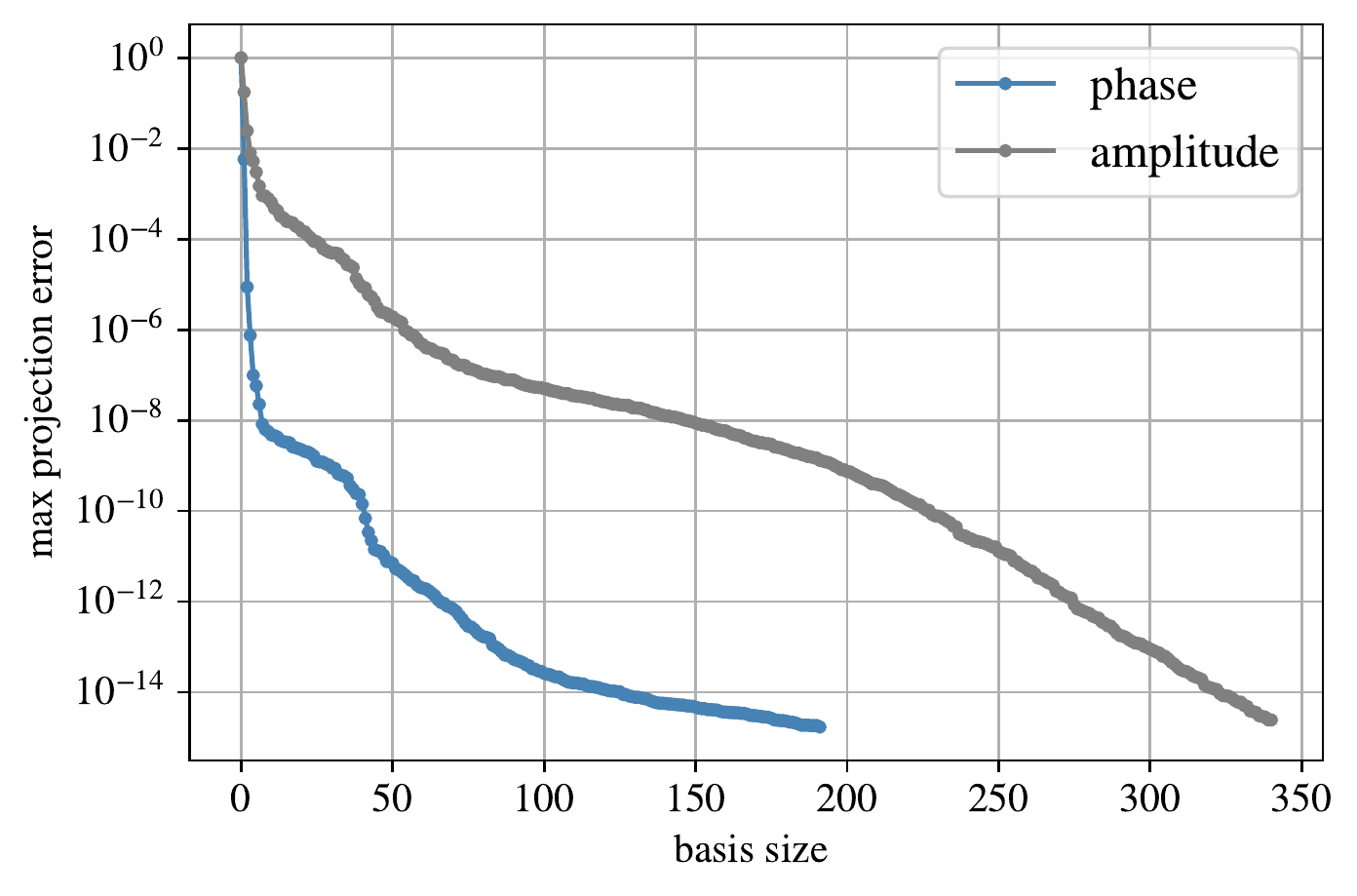}
  \caption{The greedy error as a function of the basis size. The parameters $q\in [1,5]$, $ecc\in [0,0.1]$ and $M\in[40M_\odot,100M_\odot]$.}
  \label{basis0_01nospin}
 
\end{figure}
\begin{figure}[htb]
  \centering 
  \subfigure{ 
    \includegraphics[height=2.1in]{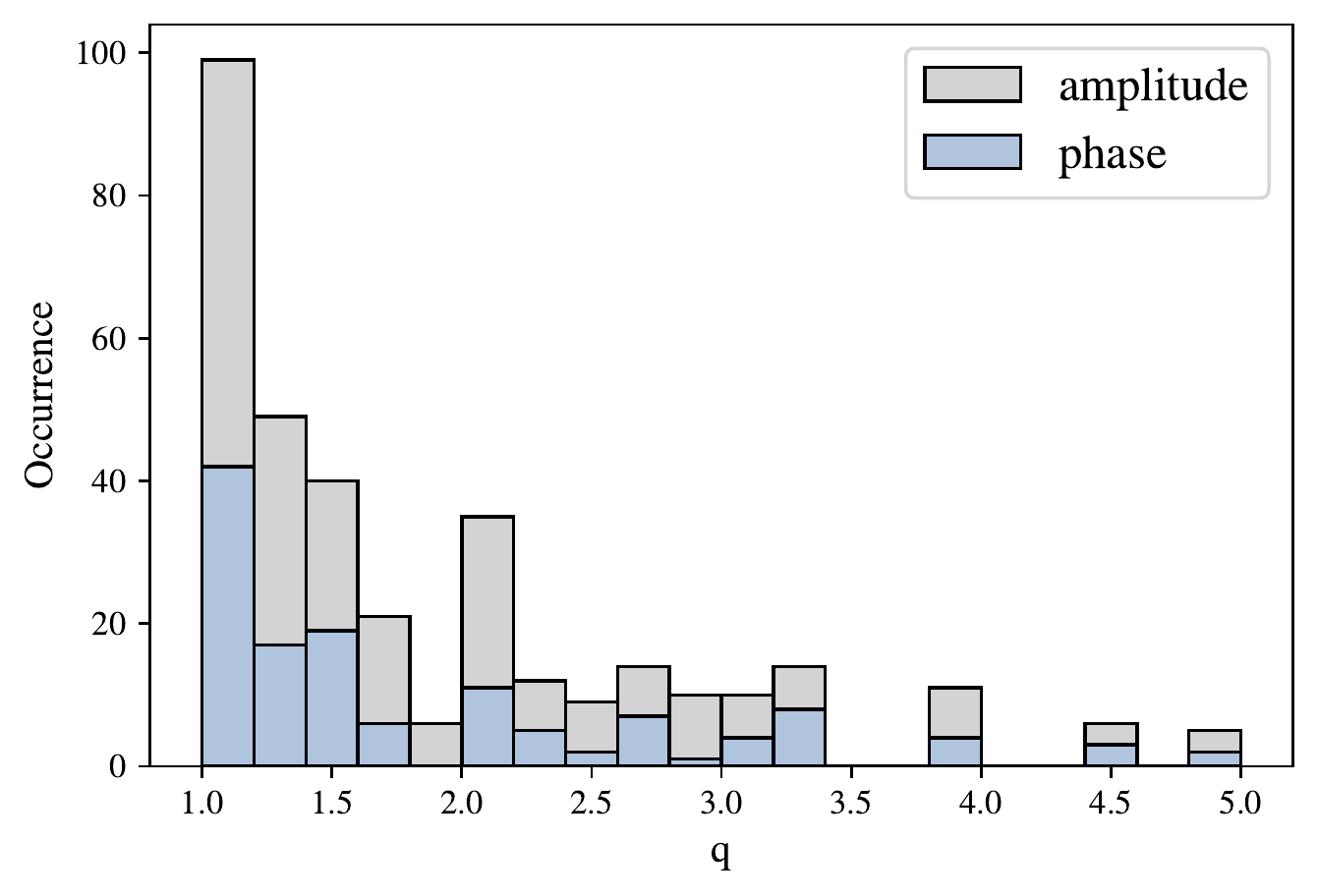} 
  } 
  \subfigure{ 
    \includegraphics[height=2.1in]{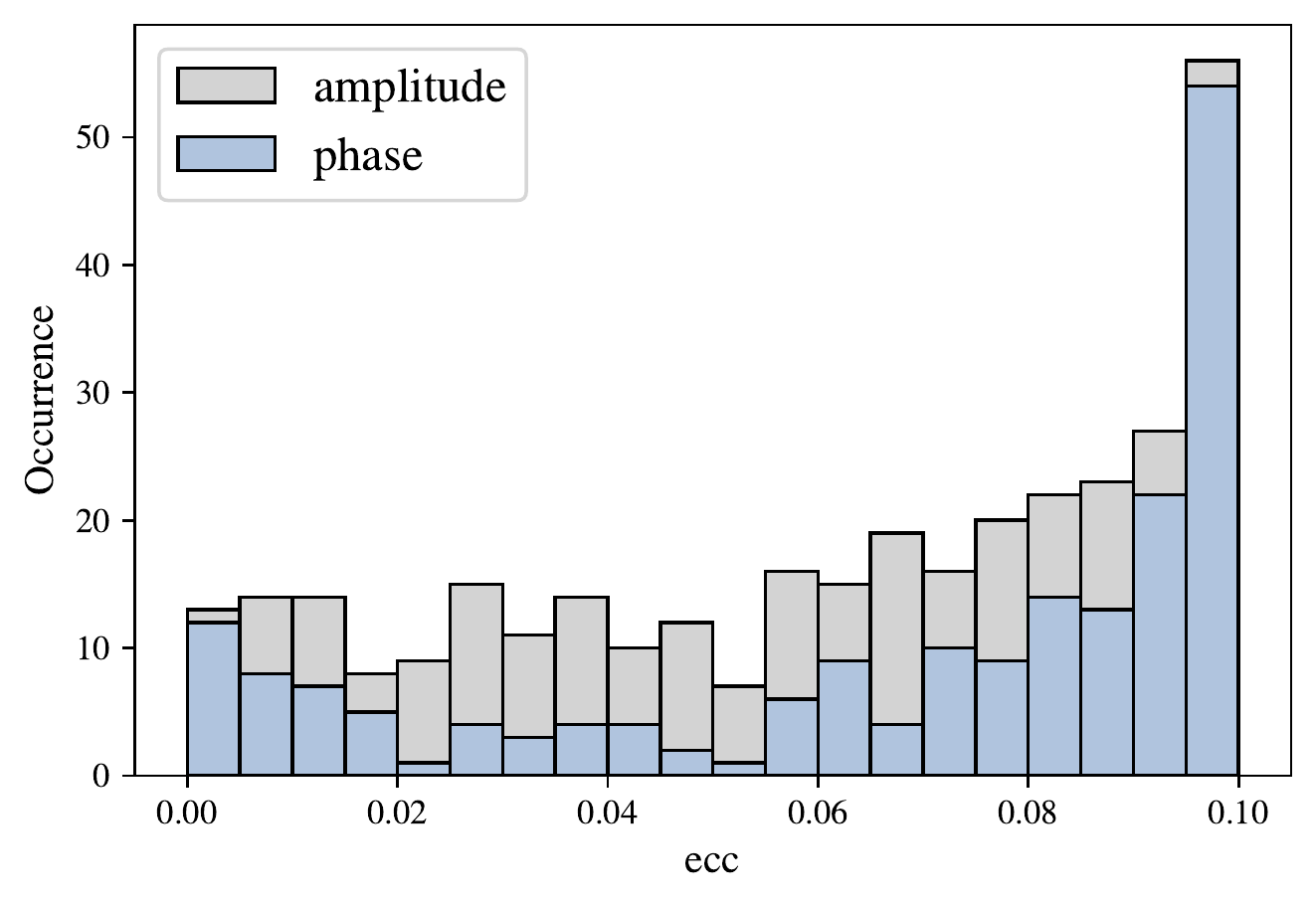} 
  } 
  
  \caption{Histogram of parameters selected by the greedy algorithm for the reduced basis of Fig.~\ref{basis0_01nospin}. }
  \label{distributionecc0_01nospin} 
\end{figure}


\begin{figure}
  \centering 
  \subfigure{ 
    \includegraphics[height=2.4in]{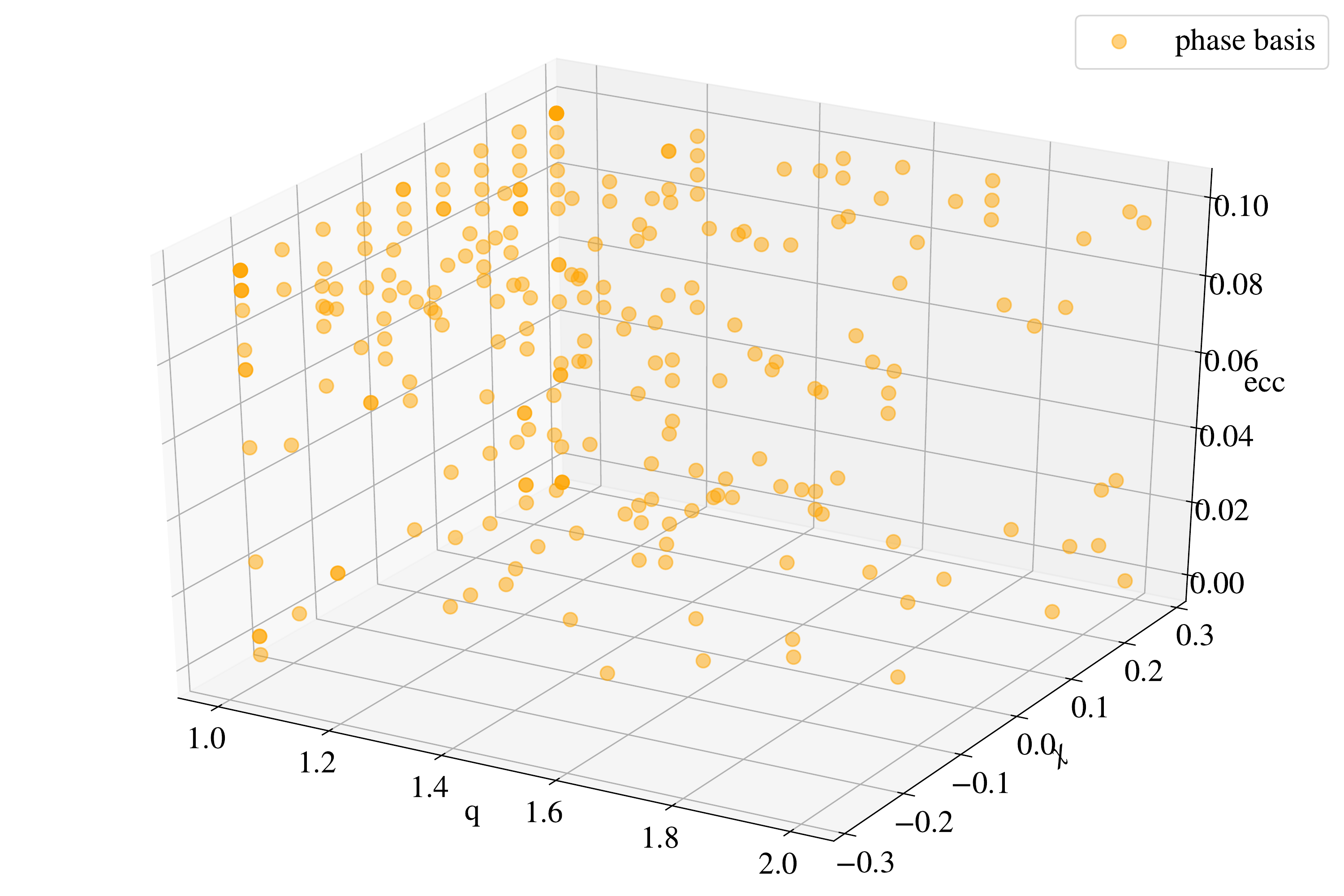}

  } 
   \subfigure{ 
    \includegraphics[height=2.4in]{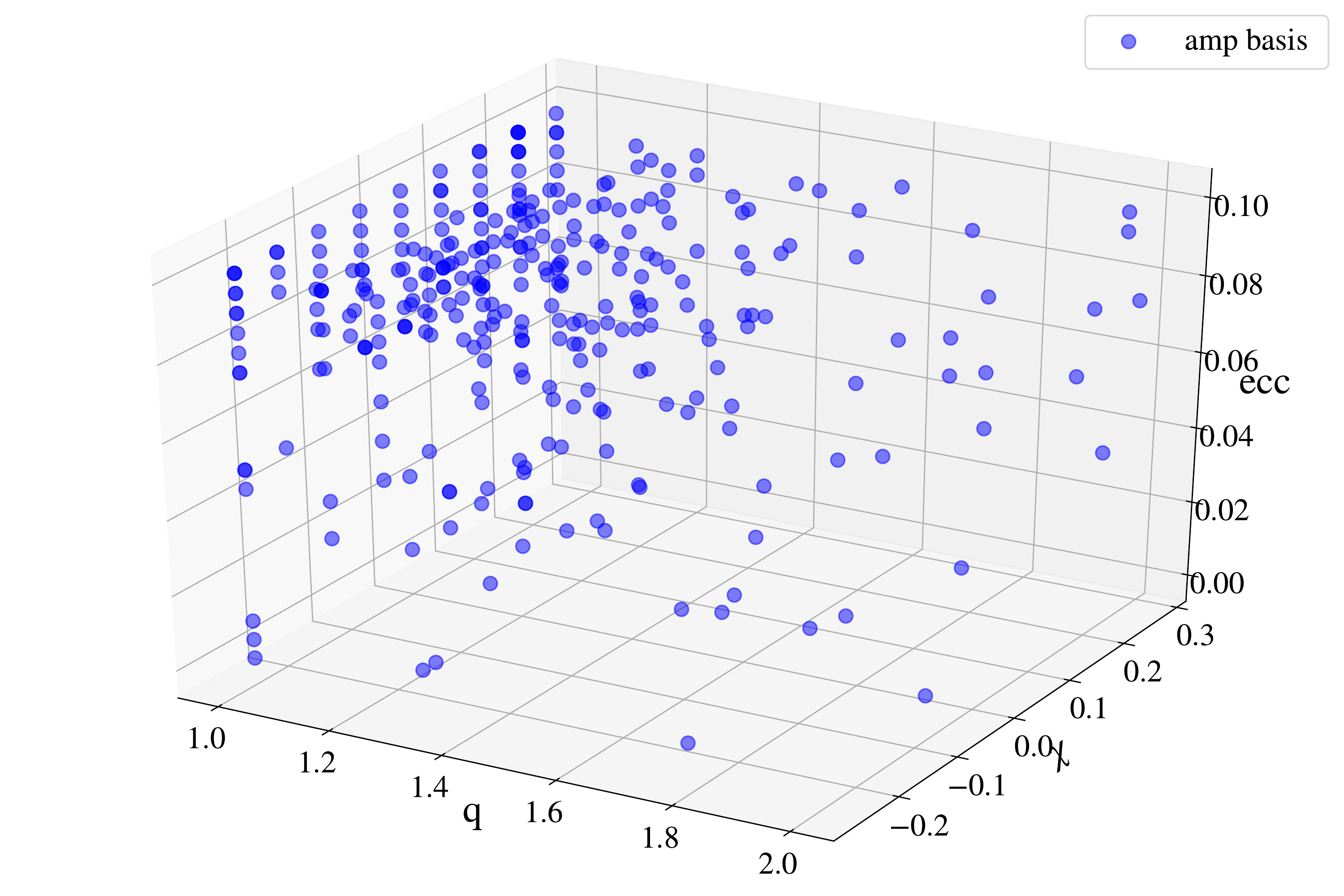}   
   } 
  \caption{The distribution for the selected phases (top panel) and amplitudes (bottom panel) with the parameters $q\in [1,2]$, $ecc\in [0,0.1]$ and $M\in[40M_\odot,100M_\odot]$. } 
  \label{selecteddata} 
\end{figure}

To built the surrogate model, we should generate a set of SEOBNRE waveforms covering each parameter space with starting frequency of 10~Hz and a sampling frequency of 4096~Hz. Therefore, we only use  the last second of each waveform. In this way, we make sure that each signal has the same length. Additionally, we consider the variation of amplitude and phase with the eccentricity separately. This makes the question easier.  We extract the phase $\phi$ and the amplitude $A$ of each time-domain waveform using $h(\lambda)=A(\lambda) e^{-i\phi(\lambda)}$ and perform the following operations:

(1) We use the reduced basis to select the most relevant $m_\phi$ and $m_A$ points from the phase and amplitude spaces respectively. The selected phases and amplitudes represent main information of the waveforms in the entire parameter space, so that we can use the smallest amount of points as the basis to achieve the required accuracy. To determine the numbers $m_\phi$ and $m_A$, we use the greedy algorithm presented in Ref.~\cite{Field:2013cfa}. By using the reduced basis  algorithm, every phase and amplitude in the original training set is approximated by an expansion of the form $ {\phi}(t ; \boldsymbol{\lambda}) \approx \sum_{i=1}^{m_{\phi}} c_{{\phi}i}(\boldsymbol{\lambda}) e_{{\phi}i}(t)$ and $ A(t ; \boldsymbol{\lambda}) \approx \sum_{i=1}^{m_A} c_{Ai}(\boldsymbol{\lambda}) e_{Ai}(t)$, where $c_{{\phi}i}$ and $c_{Ai}$ are the coefficients obtained by orthogonal projection onto the span of the basis. If the training set is dense enough, the phase and amplitude of any waveform in the training space can be well approximated. The numbers of selected points ($m_\phi$ and $m_A$) are determined by the greedy error $\sigma$ [defined in Ref.~\cite{Field:2013cfa}, Eq.~(12)]. 

In Fig.~\ref{basis0_01nospin}, we plot the greedy errors as a function of the basis sizes of phases and amplitudes in the eccentricity interval $[0-0.1]$ for nonspinning cases. From the figure, we can see how the greedy error decreases as the number of basis sizes increases. Furthermore, the phase $\phi$ (blue line) requires less basis than the amplitude $A$ (gray line). This may be because the amplitude is obviously modulated by the eccentricity.  If we set the greedy error to be less than $10^{-12}$ (following ~\cite{Field:2013cfa} and our practice proves this threshold performs very good), 341 amplitudes and 192 phases should be selected from 3366 SEOBNRE waveforms.

Fig.~\ref{distributionecc0_01nospin} shows how the selected phase and amplitude distribute when the mass ratio $q\in[1,5]$ and the eccentricity $ecc \in[0,0.1]$. When $q=1$, the numbers of selected phases and amplitudes are largest (about $40$ and $100$, respectively). As shown in the top panel of Fig.~\ref{distributionecc0_01nospin}, when the mass ratio $q$ increases, the numbers of selected phases and amplitudes have a tendency to decrease. On the other hand, when we consider the eccentricity, we can also see a tendency: the numbers of selected phases and amplitudes tend to increase as the eccentricity grows (see the bottom panel in Fig.~\ref{distributionecc0_01nospin}). The numbers of selected phases and amplitudes are largest at $ecc=0.1$ (about $55$ in both cases). In the other segments, behaviors are similar. This is because the mass ratio does not change the key feature of waveforms, but the eccentricity makes the waveforms more complicated. Therefore, the selected points must increase for the eccentric orbits to ensure the accuracy. In Table.~II, we show the numbers of selected phases and amplitudes for varied eccentricity intervals.

\begin{table}[htb]
\centering
  
  \begin{tabular}{|c|c|c|}
  \hline\hline
   ecc range   & Amplitude & Phase \\
  \hline
  $0-0.1$ & $341$  &$192$  \\
  
  $0.1-0.15$ & $367 $ & $247$  \\
  
  $0.15-0.2$ & $403 $ & $292$ \\
  
  $0.2-0.225$ & $342 $ & $250$ \\
  
  $0.225-0.25$ & $341 $ & $265$  \\
  \hline\hline
  \end{tabular}
  \caption{The numbers of the phase and amplitude we selected as the basis in each interval.}
\end{table}

For the spin-aligned case, we selected 319 basis amplitudes and 247 basis phases. In Fig.~\ref{selecteddata}, we can observe the distribution of phases and amplitudes in the three-dimensional parameter space $(q, ~\chi, ~ecc)$. We can find that the basis amplitudes and phases gather at $q\in[1,1.2]$ and $ecc\geq0.05$. This implies that more basis waveforms are needed to construct the accurate surrogate model in this region.  Similar behavior has been found in the nonspinning case.



(2)
We further process the selected phases and amplitudes to reduce the size of data. This step is significantly important because it reduces the cost for generating surrogate models. A reduced base $e_{\phi i}(t)$ or $e_{Ai}(t)$ is composed by $m_A$ fiducial amplitudes and $m_{\phi}$ fiducial phases at a certain time duration $T_i$. To find the best value of $T_i$, we build an interpolant by the empirical interpolation method. After the interpolation of the phases and amplitudes, we can describe them as $I_{mA}(t;\lambda )$ and $I_{m\phi}(t;\lambda )$ respectively. Since $I_{m\mathcal{A}}(t;\lambda)=\sum_{j=1}^{m_{A}} B_{{A}j}(\boldsymbol{\lambda}) A\left(T_{Aj} ; \boldsymbol{\lambda}\right)$ and $I_{m\mathcal{\phi}}(t;\lambda)=\sum_{j=1}^{m_{\phi}} B_{{\phi}j}(\boldsymbol{\lambda}) {\phi}\left(T_{{\phi}j} ; \boldsymbol{\lambda}\right)$, the coefficients $\left\{B_{Ai}\right\}_{i=1}^{m}$ and $\left\{B_{{\phi}i}\right\}_{i=1}^{m}$ can be defined by using the matrices $V_{A},~V_{\phi}$, and the reduced basis of amplitude $e_{Ai}(t)$ and phase $e_{{\phi}i}(t)$, where the interpolation matrices are
\begin{equation}
V_{A} \equiv\left(\begin{array}{cccc}
e_{A1}\left(T_{A1}\right) & e_{A2}\left(T_{A1}\right) & \cdots & e_{Am}\left(T_{A1}\right) \\
e_{A1}\left(T_{A2}\right) & e_{A2}\left(T_{A2}\right) & \cdots & e_{Am}\left(T_{A2}\right) \\
e_{A1}\left(T_{A3}\right) & e_{A2}\left(T_{A3}\right) & \cdots & e_{Am}\left(T_{A3}\right) \\
\vdots & \vdots & \ddots & \vdots \\
e_{A1}\left(T_{Am}\right) & e_{A2}\left(T_{Am}\right) & \cdots & e_{Am}\left(T_{Am}\right)
\end{array}\right),
\end{equation}
and
\begin{equation}
V_{\phi} \equiv\left(\begin{array}{cccc}
e_{{\phi}1}\left(T_{{\phi}1}\right) & e_{{\phi}2}\left(T_{{\phi}1}\right) & \cdots & e_{{\phi}m}\left(T_{{\phi}1}\right) \\
e_{{\phi}1}\left(T_{{\phi}2}\right) & e_{{\phi}2}\left(T_{{\phi}2}\right) & \cdots & e_{{\phi}m}\left(T_{{\phi}2}\right) \\
e_{{\phi}1}\left(T_{{\phi}3}\right) & e_{{\phi}2}\left(T_{{\phi}3}\right) & \cdots & e_{{\phi}m}\left(T_{{\phi}3}\right) \\
\vdots & \vdots & \ddots & \vdots \\
e_{{\phi}1}\left(T_{{\phi}m}\right) & e_{{\phi}2}\left(T_{{\phi}m}\right) & \cdots & e_{{\phi}m}\left(T_{{\phi}m}\right)
\end{array}\right).
\end{equation}

Therefore the coefficients can be written as
\begin{equation}
    B_{Aj}(t) \equiv \sum_{i=1}^{m_{A}} e_{Ai}(t)\left(V_{A}^{-1}\right)_{i j}
\end{equation}
and 
\begin{equation}
    B_{{\phi}j}(t) \equiv \sum_{i=1}^{m_{\phi}} e_{{\phi}i}(t)\left(V_{{\phi}}^{-1}\right)_{i j},
\end{equation}
where $A\left(T_{Aj};\boldsymbol{\lambda}\right)$ and ${
\phi}\left(T_{{\phi}j};\boldsymbol{\lambda}\right)$ are the basis of amplitudes and the phases at empirical nodes respectively.

(3) In this step, by using the data $A\left(T_{Aj};\boldsymbol{\lambda}\right)$ and ${
\phi}\left(T_{{\phi}j};\boldsymbol{\lambda}\right)$, we fit out polynomials to predict the waveforms. We employ least squares to fit the amplitudes and phases with these relations
\begin{equation}
A_{i}(\mathbf{\Lambda})=\sum_{n=0}^{\alpha_{Ai}} a_{Ai, n} \Lambda^{n}, \text{and} \quad \phi_{i}(\mathbf{\Lambda})=\sum_{n=0}^{\beta_{{\phi}i}} b_{{\phi}i, n} \Lambda^{n} \,.
\end{equation}
where $\alpha_{i}$ and $\beta_{i}$ are the degrees of the polynomials at the empirical time $T_{Ai}$ for $i=1,2, \ldots, m_{A}$ and $T_{{\phi}i}$ for $i=1,2, \ldots, m_{{\phi}}$ respectively. Both $\alpha_{i}$ and $\beta_{i}$ are less than $m$, and in this paper, we set $\alpha_{Ai}=3$ and $\beta_{{\phi i}}=5$ to obtain good performance of the surrogate model. The $\mathbf{\Lambda}$ is a three-dimensional set of parameters for the nonspinning binaries ($\mathbf{\Lambda_{nonspining}}=\left[q, M, ecc\right]$) and a five-dimensional set of parameters for the spin-aligned cases ($\mathbf{\Lambda_{spin-aligned}}=\left[q, M, \chi_{1}, \chi_{2}, ecc\right]$).

(4)  Finally, the time-domain surrogate model SEOBNRE\_S for eccentric-spinning BBHs is

\begin{equation}
\begin{array}{c}
h_{\mathrm{S}}(t ; q, M, \chi_{1}, \chi_{2}, ecc) \equiv \\ \sum_{i=1}^{m_A} B_{A i}(t) A_{i}(q, M, \chi_{1}, \chi_{2}, ecc) 
e^{-i  \sum_{i=1}^{m_{\phi}B_{{\phi} i}\phi_{i}(q, M, \chi_{1}, \chi_{2}, ecc)}}\,.
\end{array}
\end{equation}

\begin{figure*}[htb]
  \centering
  
  \subfigure[]{ 
    \includegraphics[height=2.25in]{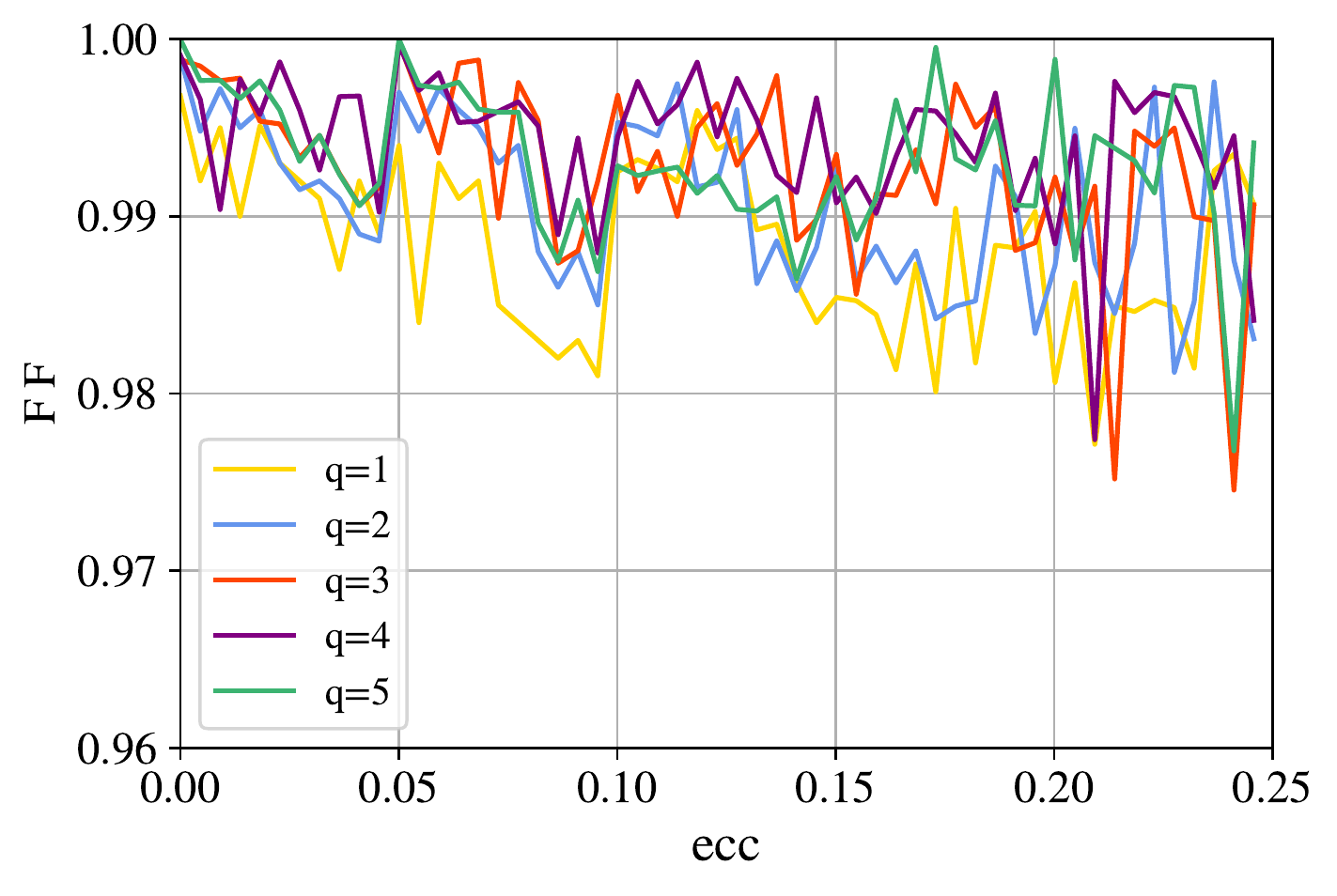} 
  } 
  \subfigure[]{ 
    \includegraphics[height=2.25in]{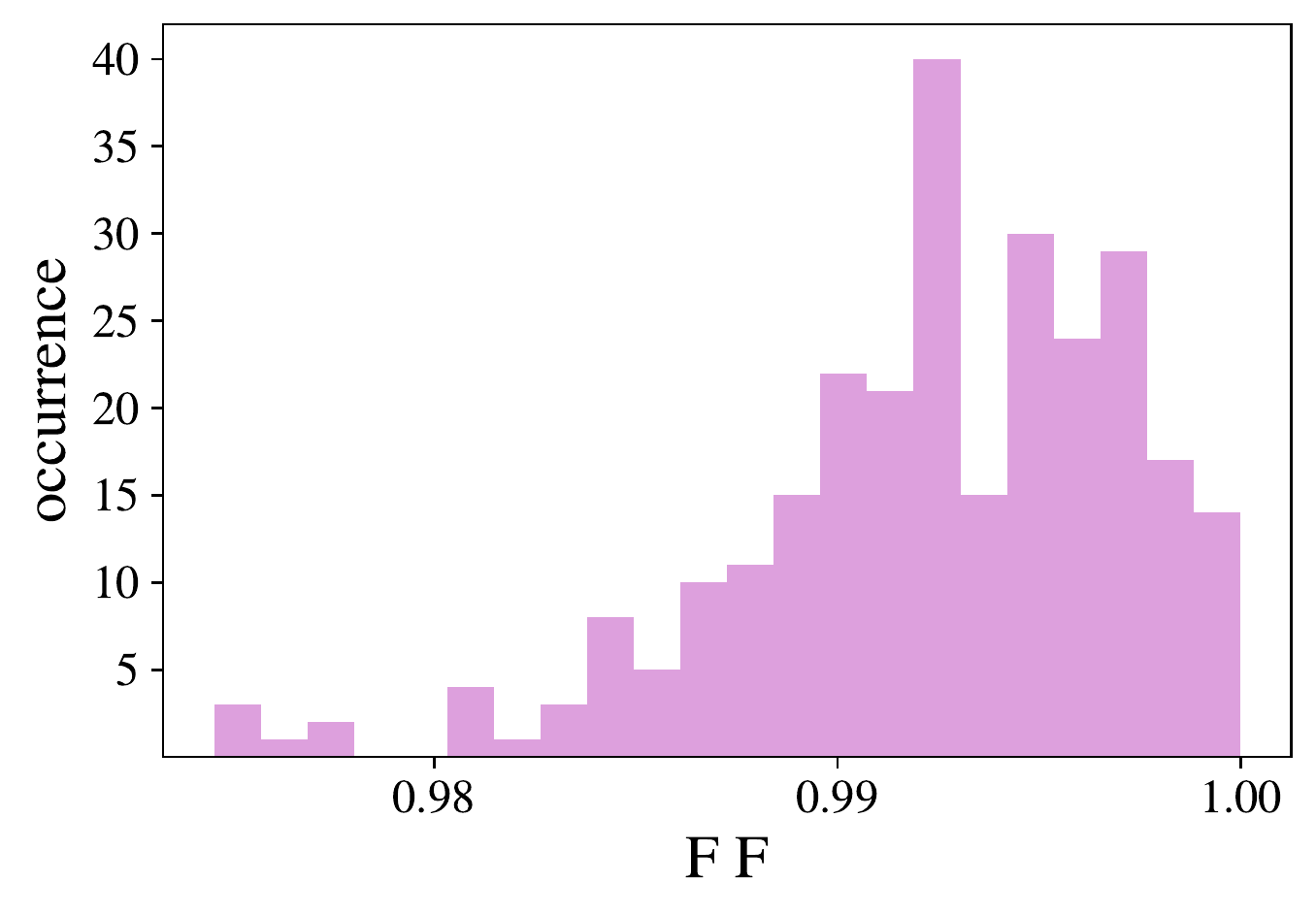} 
  } 
  
  \caption{(a) The overlap of SEOBNRE waveforms and surrogate waveforms for the nonspinning case as a function of the eccentricity($ecc$) for different values of the mass ratios. The range of the eccentricity range is [0,0.25]. (b) The count of the fitting factor (FF) for all the waveforms compared in (a). } 
  \label{eccnospinFFandcount} 
\end{figure*}
\begin{figure*}
  \centering 
  \includegraphics[height=4.6in]{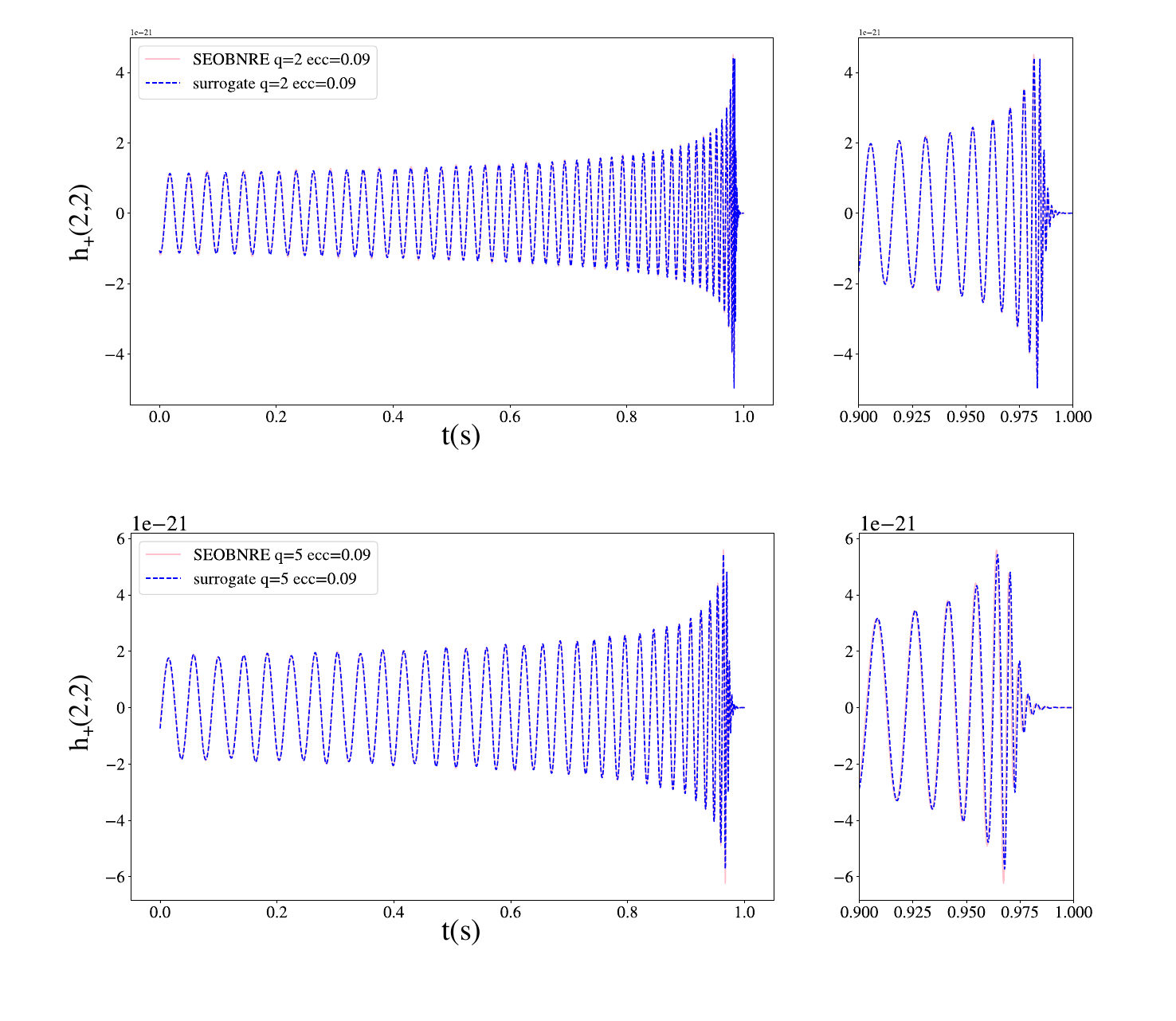}
  \caption{The last one second of SEOBNRE waveforms and surrogate waveforms for nonspinning BBH binary with $q=2, q=5$.   The  eccentricity of waveforms is $ecc =0.09$.  }
  \label{ecc009q2q5waveform}
  
\end{figure*}

\section{Accuracy and efficiency of the surrogate model} 
In this section, we first validate our surrogate model SEOBNRE\_S with the original SEOBNRE waveforms, then we compare the computation speeds of them.



\subsection{Model validation}


Since the SEOBNRE waveforms $h_{\mathrm{SEOBNRE}}$ and surrogate waveforms $h_{\mathrm{s}}$ are inspiral, merge, and ringdown waveforms, both of them have their amplitude peaks. To compare the waveforms, we need to align the two amplitude peaks at $t=0$ and make sure they have the same length. We then compare the two waveforms using their inner product weighted by the power spectral density of the detector noise $S_n(f)$
\begin{align}\label{equ:inner}
    \left\langle h_{\rm SEOBNRE},  h_{\mathrm{s}}\right\rangle= 4 \operatorname{Re} \int_{f_{\rm min }}^{f_{\rm max }} \frac{\tilde{h}_{\rm SEOBNRE}(f) \tilde{h}_{\mathrm{s}}^{*}(f)}{S_{n}(f)} d f \,.
 \end{align}
 where ``$\operatorname{Re}$" means taking the real part, $f_{\max }$ corresponds to the sampling rate of waveform and $f_{\min }$ corresponds to the duration time of the waveform. $\tilde{h}$ means the Fourier transformation of the time series $h(t)$ and ``$*$" the complex conjugate. $S_n(f)$ is taken from the LIGO's sensitivity curve.
 
Based on the inner product, the fitting factor of two signals is
\begin{equation}
    {\rm FF}=\max _{t_{0}, \phi_{0}} \frac{\left(h_{\mathrm{SEOBNRE}}, h_{\mathrm{s}}\right)}{\sqrt{\left(h_{\mathrm{SEOBNRE}}, h_{\mathrm{SEOBNRE}}\right)\left(h_{\mathrm{s}}, h_{\mathrm{s}}\right)}}.
\end{equation}
Hence, the mismtach of two signals is
\begin{equation}
    {\rm Mismatch} = 1- {\rm FF}
    \label{mismatcheq}
\end{equation}

Now, we evaluate our surrogate model for the nonspinning cases. In this situation, there are three intrinsic parameters: the mass ratio, the total mass, and the orbital eccentricity. We  calculate the fitting factors for varied binary parameters and the results are shown in Fig.~\ref{eccnospinFFandcount}. We can see that most fitting factors are better than 98 \%. The statistical distribution of the FF values is shown in Fig.~\ref{eccnospinFFandcount}(b). These results demonstrate our surrogate model is quiet faithful.

\begin{figure*}
  \centering 
  \includegraphics[height=8.2in]{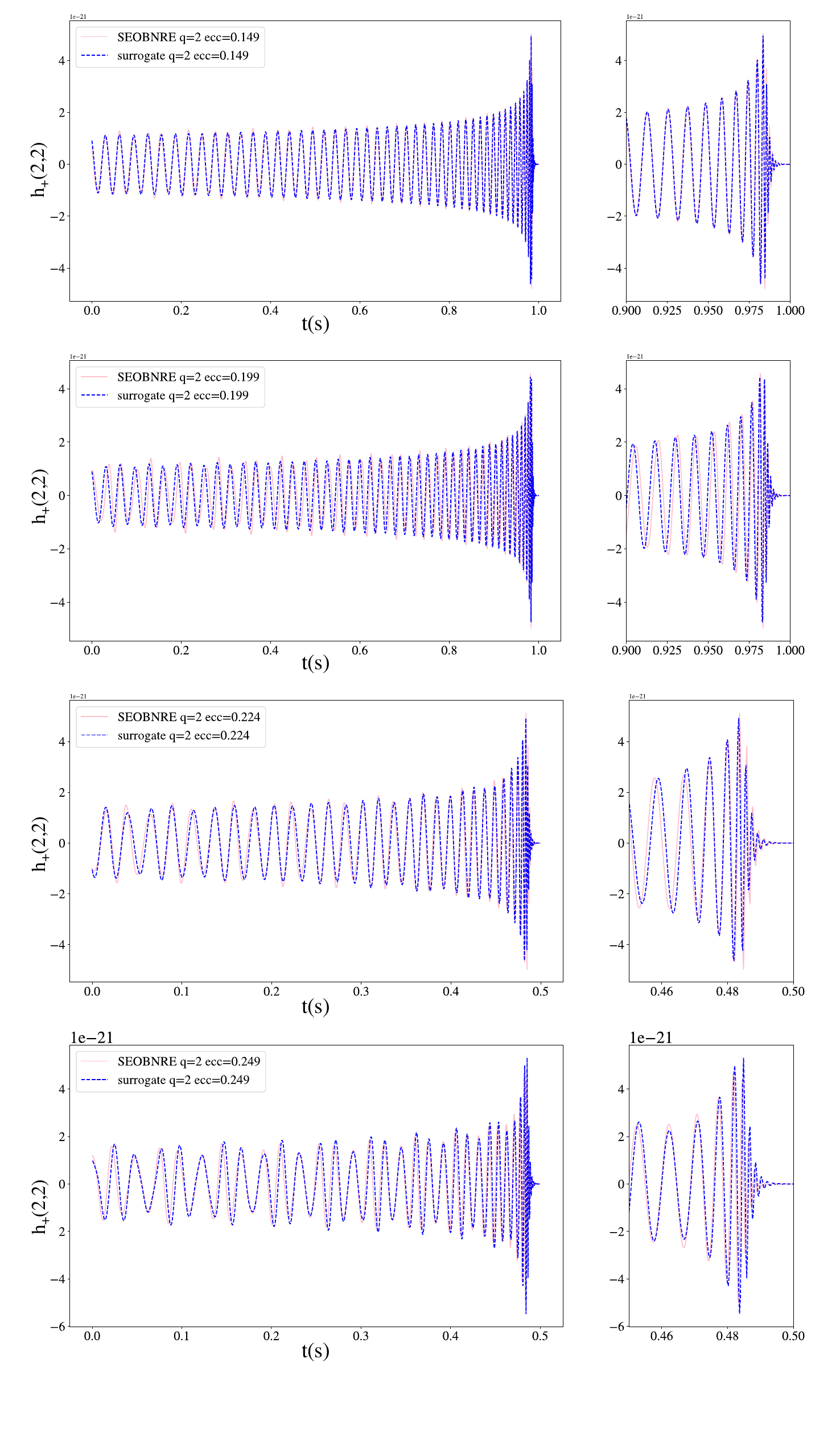}
  \caption{The last second of SEOBNRE waveforms and surrogate waveforms for nonspinning BHH with $q=2$. The  eccentricity of waveforms is $ecc =0.159,0.199,0.224$, and $0.249$ from top to bottom.}
  \label{q2waveformnospineccall}
  
\end{figure*}

In Fig.~\ref{ecc009q2q5waveform}, we plot waveforms generated by the SEOBNRE and the surrogate model with eccentricity $ ecc=0.09$, mass ratio $q=2$ (top panel) and $q = 5$ (bottom panel). The two kinds of waveforms match perfectly and the FF is $0.99$.  Moreover, for larger eccentricities, the surrogate waveforms  can still coincide the SEOBNRE waveforms very well. This can be seen in Fig.~\ref{q2waveformnospineccall}, where we compare four waveforms with $ecc=0.149,~ 0.199,~ 0.2249,~ 0.249$ and $q = 2$. The FFs are $0.988$, $0.988$,$0.987$ and $0.984$ for $ecc=0.149, 0.199, 0.2249 $ and $0.249$ respectively.

\begin{figure*}
  \centering 
  \includegraphics[height=2.1in]{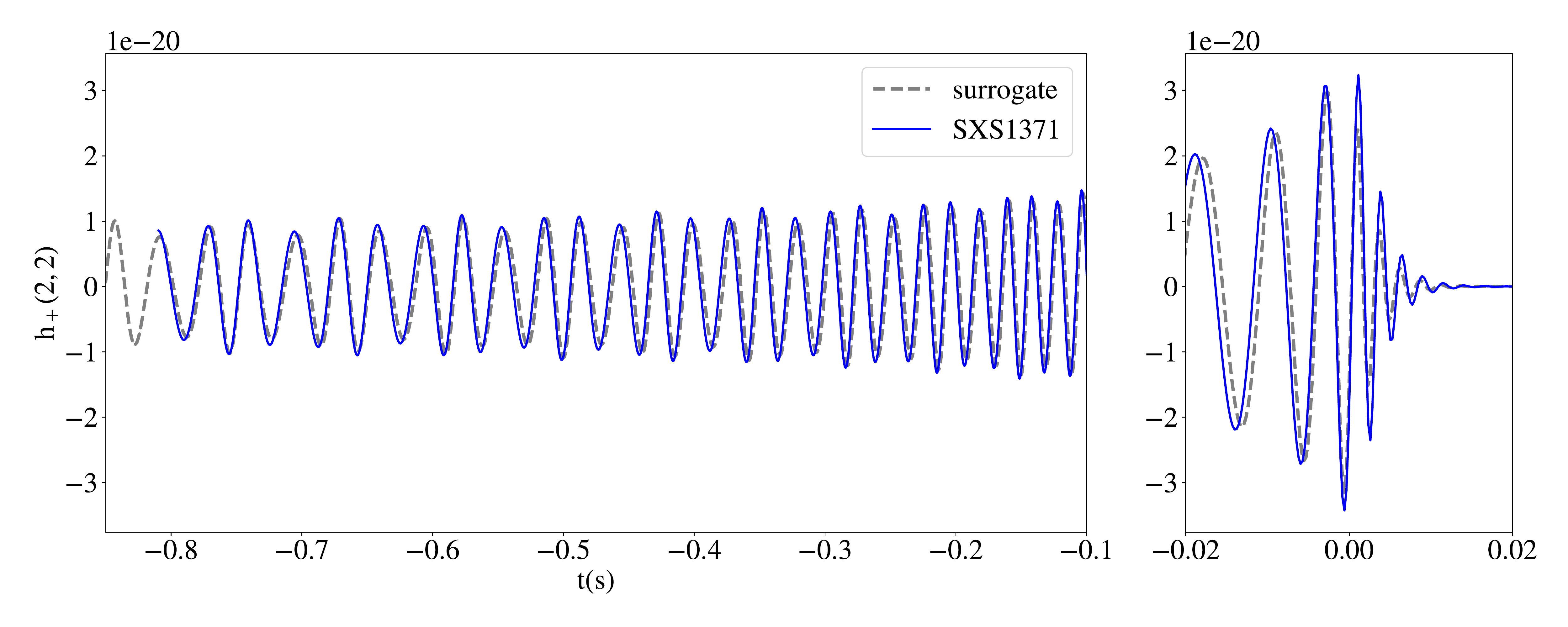}
  \caption{The surrogate waveform (dashed line) and the NR waveform SXS:BBH:1371 (blue line). The parameters for SXS: 1371 are:$q=3$, $M=40_\odot$ and $ecc=0.05$ at 26~Hz. The parameters for the surrogate model are $q=3$, $M=40_\odot$ and $ecc=0.145$ at 10~Hz.}
  \label{sxs}
  
\end{figure*}

In addition, we compare our surrogate waveform with the SXS waveform. Recently, Ref.~\cite{Islam:2021mha} has proposed a surrogate model for the eccentric BBHs without spin. They compared their waveform with the NR waveform SXS: BBH:1371, and the overlap is about $0.99$. The parameters for SXS: 1371 are:$q=3$, $M=40_\odot$ and $ecc=0.05$ at 26~Hz. The parameters for this surrogate waveform are: $q = 3$, $M = 40M\odot$, and $ecc = 0.050$.  In Fig.~\ref{sxs}, we  plot our surrogate waveform with the NR waveform SXS: BBH:1371. The fitting factor between the two waveforms is 0.9903, which implies that our surrogate model can coincide with the one in ~\cite{Islam:2021mha} for the nonspinning cases. The parameters for surrogate model are: $q=3$, $M=40_\odot$ and $ecc=0.145$ at 10~Hz. We also compare the original SEOBNRE waveform with the SXS: BBH:1371, and the overlap is $0.9909$. The parameters for SEOBNRE are the same as SEOBNRE\_S.

\begin{figure}[hb]
  \centering 
  \subfigure{ 
    \includegraphics[height=2.1in]{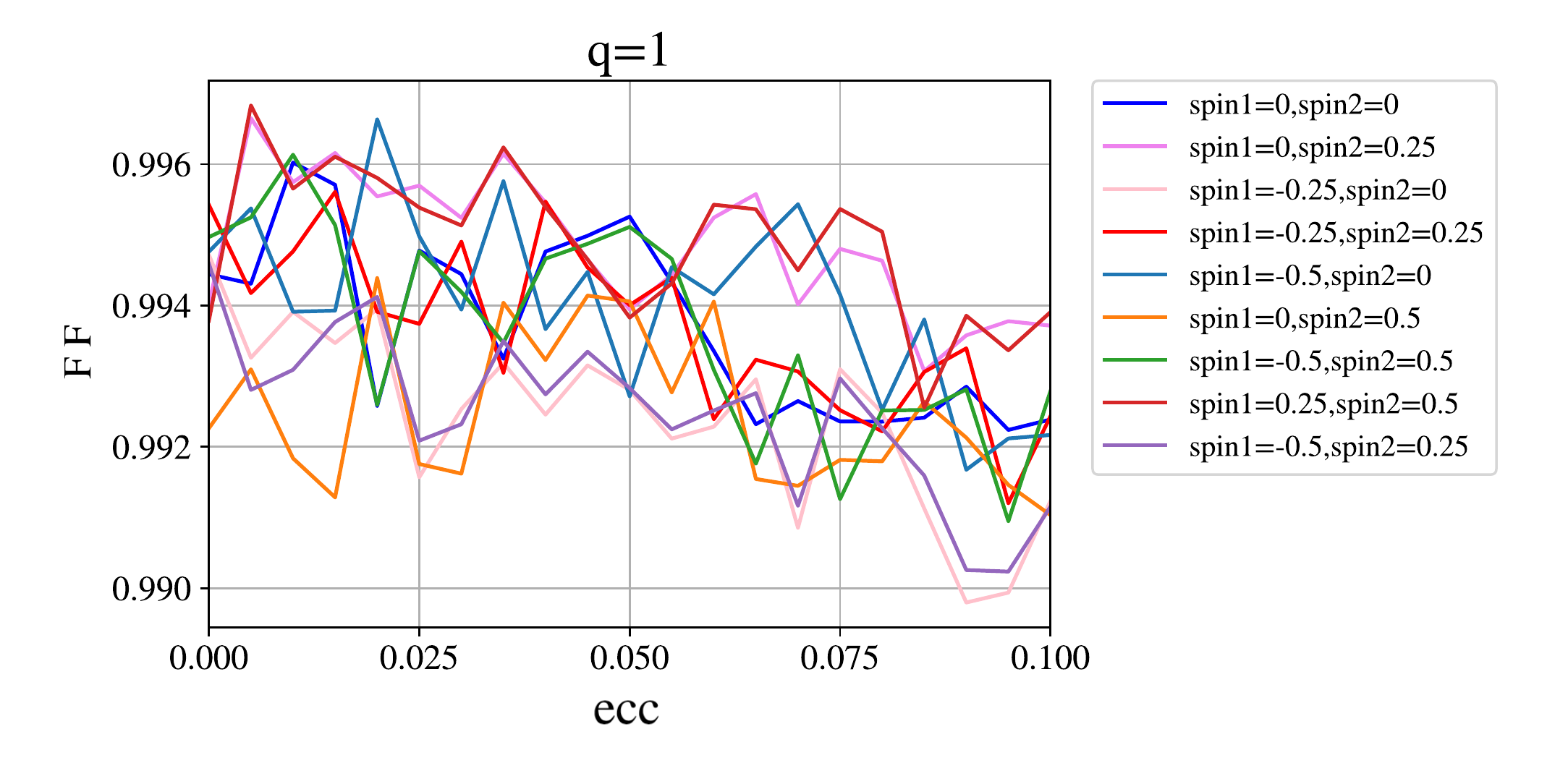}
    
  } 
  \subfigure{ 
    \includegraphics[height=2.1in]{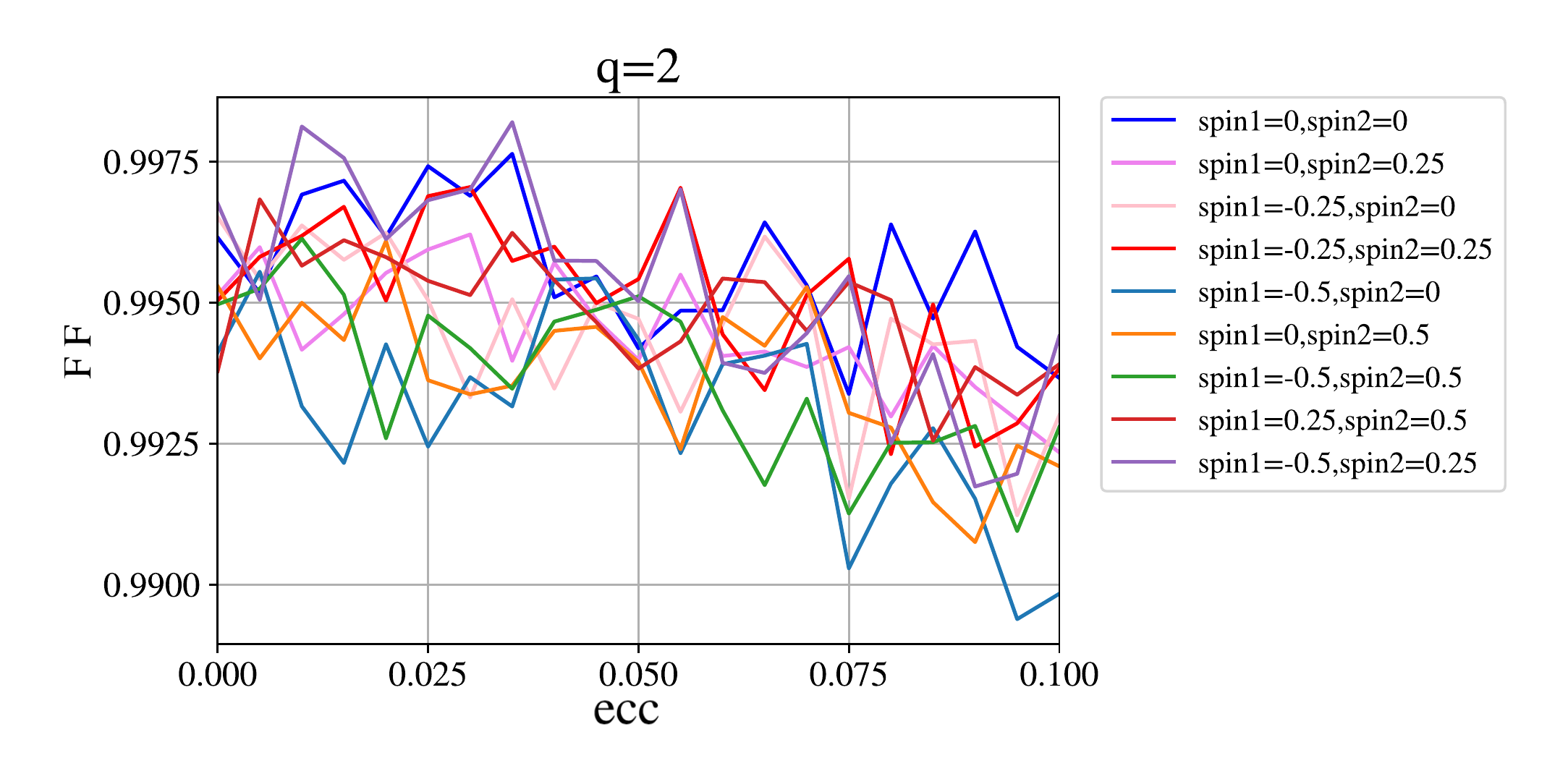}
    
  } 
    
 
  \caption{The fitting factor (FF) as a function of $ecc$ for the surrogate waveform and the SEOBNRE waveform for different values of the mass ratio $q=1$ (top panel) and $q=2$ (bottom panel)}
  \label{spin-alignedff} 
\end{figure}

\begin{figure}
  \centering 
  \includegraphics[height=2.3in]{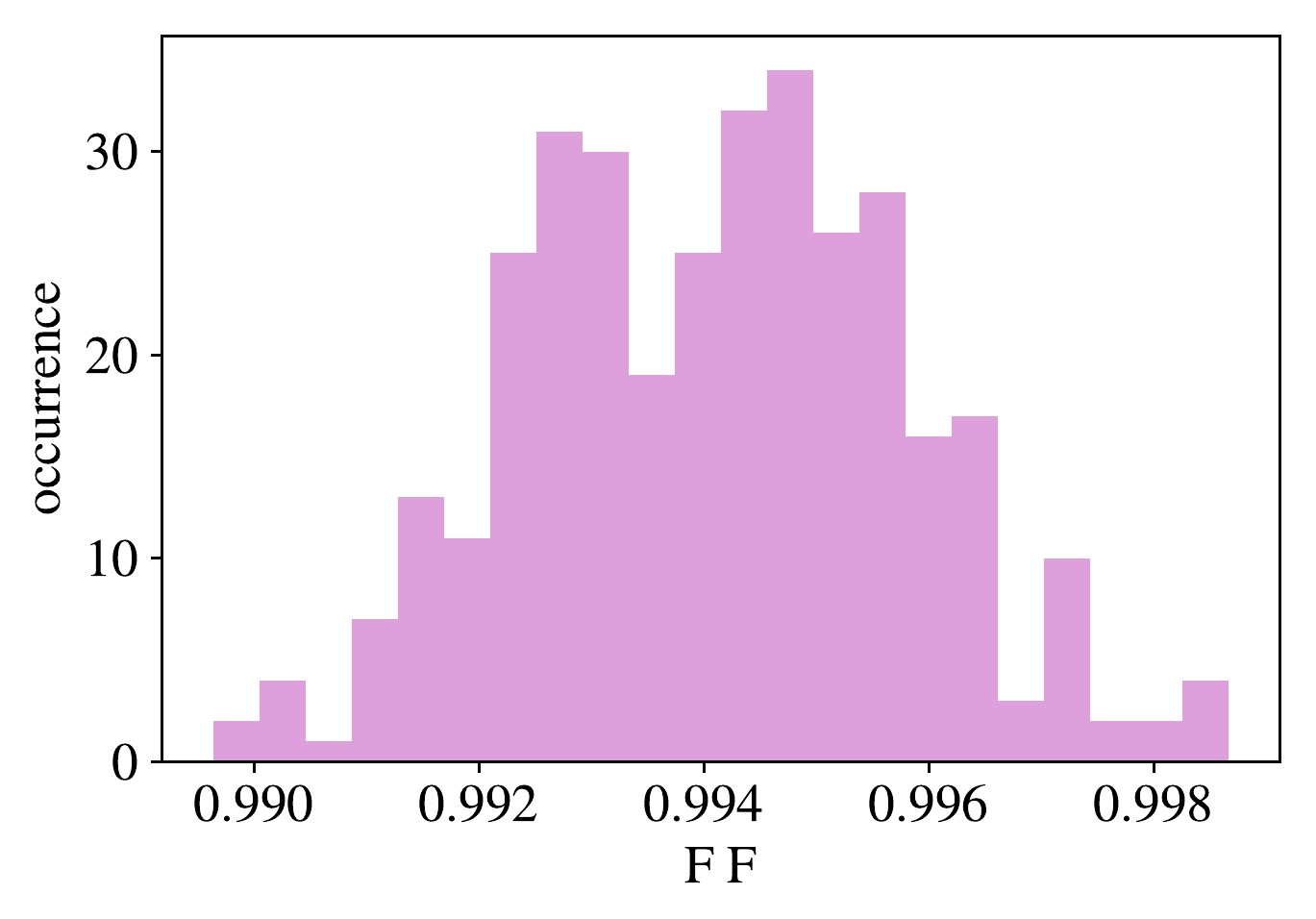}
  \caption{The count of the FF for all the waveforms compared in Fig. \ref{spin-alignedff}.}
  \label{spinffcount}
  
\end{figure}

For the spin-aligned cases, there are five intrinsic parameters: the mass ratio $q$, the total mass $M$, two dimensionless spins $\chi_1$ and $\chi_2$, and the orbital eccentricity $ecc$. In Fig.~\ref{spin-alignedff}, the FF values are displayed for $q=1$ (top panel) and $q=2$ (bottom panel). It shows that the surrogate waveforms match the SEOBNRE ones very well for the given spins. From the figure, it is also possible to see that the FF decreases as the eccentricity increases. For example, when the eccentricity is close to 0.1, a few values drop below 99\%. However, according to the statistical chart shown in Fig.~\ref{spinffcount}, we find that
most of the FF values are better than 99\%.

\begin{figure*}
  \centering 
  { 
    \includegraphics[height=4.6in]{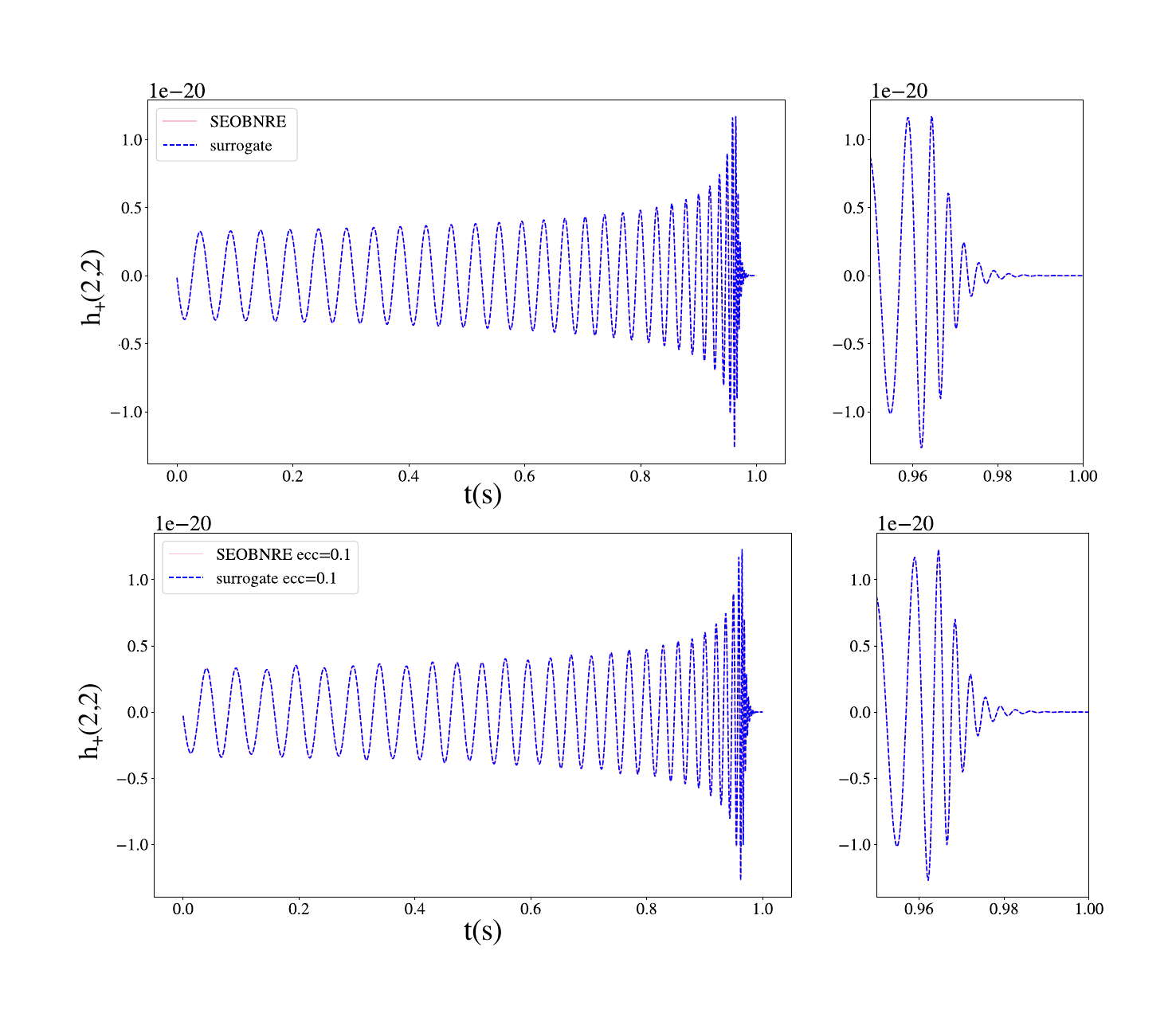}
   
  } 
 
  \caption{The last second of SEOBNRE waveforms and surrogate waveforms for the circular(top panle) and eccentric (bottom panel) BBHs with $m_1=36$, $m_2=29$, $\chi_1=0.33$, $\chi_2=-0.44$. } 
  \label{waveformwithspin} 
\end{figure*}

In Fig.~\ref{waveformwithspin}, we plot two waveforms to compare our surrogate model with the SEOBNRE one. In the top panel of Fig.~\ref{waveformwithspin}, we use the same parameters of GW150914, i.e. $m_1=36$, $m_2=29$, $\chi_1=0.33$, $\chi_2=-0.44$, and take $ecc=0$. In the bottom panel of Fig.~\ref{waveformwithspin}, we take into account the eccentricity ($ecc=0.09$). In both panels, we can see that the waveforms from the surrogate and SEOBNRE models match quite well with fitting factors $0.993$ and $0.995$ respectively. All these comparisons demonstrate that our surrogate model SEOBNRE\_S performs very well for the eccentric BBHs in a range of mass ratio with aligned spins. 





\subsection{Generation time for waveforms}
\begin{figure}
  \centering 
  \includegraphics[height=2.1in]{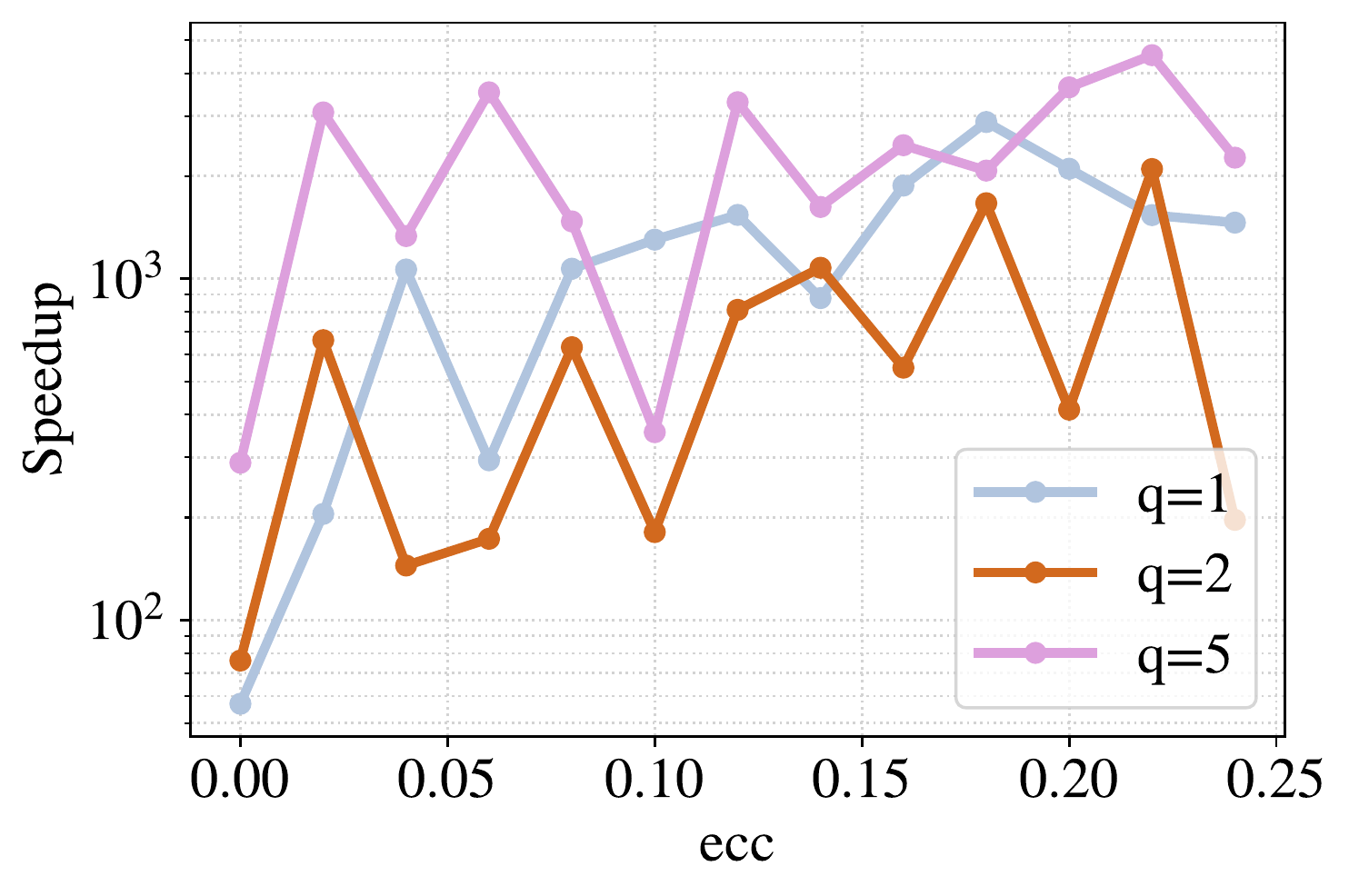}
  \caption{
  The speedup of the surrogate model versus the SEOBNRE one for varied mass ratios. The speedup is defined by the ratio of generation times of the SEOBNRE and SEOBNRE\_S for the same BBH,  and the starting frequency is  $Mf=10^{-3}$.}
  \label{speedup}
  
\end{figure}
In the previous section, we have validated the accuracy of our surrogate model. Now, we test the efficiency of this surrogate model versus the original SEOBNRE one. 

We compute the SEOBNRE and surrogate waveforms for the same BBH parameters with a single core (Intel(R) CPU E5-2687W 0 @ 3.10Ghz) and count their generation times. To ensure that the length of the waveforms for different $M$ is equal, we set the starting frequency of waveforms as $Mf=10^{-3}$. By comparing the generation times, we calculate the speedup of the surrogate model and the results are shown in Fig ~\ref{speedup}. It shows that the surrogate model can achieve a good speedup compared to the SEOBNRE model. 

Meanwhile, the speedup effect tends to be better for the higher eccentricity. This may be due to the computation cost of SEOBNRE model will increase for large eccentricities, but the generation time of the surrogate model has no obvious change for varied eccentricities. When $q=5$, the speedup of the surrogate model versus the SEOBNRE model is remarkable (about $10^{3}$ times faster) because the generation time of an SEOBNRE waveform is much longer than the generation time in the other two cases ($q=1$ and $q=2$) due to the more waveform cycles. From this figure, the surrogate model is about $10^2-10^{3}$ times faster than the SEOBNRE. Therefore, the time to generate $10^6$ waveforms with eccentricities reduces from several years to about one single day by using one CPU core. This improvement enables our surrogate model to be used to analyze the GW data and do parameter estimations for the potential eccentric BBHs. We are planning to reanalyze the GW190521 event with our SEOBNRE\_S in the next work.

\section{Conclusion and outlook}
In the present paper, we construct a surrogate model (SEOBNRE\_S) for eccentric binary black hole mergers based on the SEOBNRE waveforms and the reduced order model (Rompy). Our SEOBNRE\_S model can generate eccentric waveforms up to $e = 0.25$ with parallel spins up to $\chi = 0.5$, and the speed of generation is up to a few thousand faster than the original SEOBNRE model with mismatch $\lesssim 0.01$. The mass ratio in our surrogate model can be as large as 1:5.  

This stupendous acceleration makes SEOBNRE\_S  an appropriate template for GW data analysis to find the potential eccentricity in the binary mergers. We compare the SEOBNRE\_S waveforms with SEOBNRE ones in detail and assure that both models coincide very well with eccentricity less than 0.25, spin less than 0.5, and mass ratio not exceeding 1:5. Due to the accuracy of SEOBNRE has been confirmed in several works \cite{liu2019validating, Yun_2020}, we can believe that the new surrogate model also has enough accuracy in the above parameter space. However, more comprehensive comparisons with other waveforms, including other surrogate models, are also helpful. For example, we compare our model with the nonspinning surrogate model NRSur2dq1Ecc~\cite{Islam:2021mha}, and find both of them coincide very well the numerical relativity waveform for the nonspinning case. Furthermore, data analysis with our model for some potential eccentric BBH candidates in the LIGO-Virgo catalogs should be attractive. We will leave these for the subsequent work.

SEOBNRE\_S now is only valid for the eccentricity up to 0.25. Though this may cover most cases of eccentrics, it is still valuable to extend to higher eccentricity. It is challenging due to the much more complicated waveforms when $ecc$ is wild. The surrogate model for a wide range of eccentricity will be our next task in the future.   

\section*{ACKNOWLEDGMENTS} This work is supported by NSFC No. 11773059. MEXT also supported this work, JSPS Leading-edge Research Infrastructure Program, JSPS Grant-in-Aid for Specially Promoted Research 26000005, JSPS Grant-in-Aid for Scientific Research on Innovative Areas 2905: JP17H06358, JP17H06361, and JP17H06364, JSPS Core-to-Core Program A. Advanced Research Networks, JSPS Grant-in-Aid for Scientific Research (S) 17H06133, the joint research program of the Institute for Cosmic Ray Research, University of Tokyo, and Key Research Program of Frontier Sciences, CAS, No. QYZDB-SSW-SYS016.

\newpage


\end{document}